\input harvmac
\input amssym
\input epsf
\let\includefigures=\iftrue
%
% the following is to use blackboard bold fonts --
%\let\useblackboard=\iftrue
%
% activate this if you don't have them.
%\let\useblackboard=\iffalse
%
% You might also need to remove this line.
\newfam\black
\noblackbox
\includefigures
\message{If you do not have epsf.tex (to include figures),}
\message{change the option at the top of the tex file.}
\def\figin{\epsfcheck\figin}\def\figins{\epsfcheck\figins}
\def\epsfcheck{\ifx\epsfbox\UnDeFiNeD
\message{(NO epsf.tex, FIGURES WILL BE IGNORED)}
\gdef\figin##1{\vskip2in}\gdef\figins##1{\hskip.5in}% blank space instead
\else\message{(FIGURES WILL BE INCLUDED)}%
\gdef\figin##1{##1}\gdef\figins##1{##1}\fi}
\def\DefWarn#1{}

\def\figinsert{\goodbreak\midinsert}
\def\ifig#1#2#3{\DefWarn#1\xdef#1{fig.~\the\figno}
\writedef{#1\leftbracket fig.\noexpand~\the\figno}%
\figinsert\figin{\centerline{#3}}\medskip\centerline{\vbox{\baselineskip12pt
\advance\hsize by -1truein\noindent\footnotefont{\bf
Fig.~\the\figno:} #2}}
\bigskip\endinsert\global\advance\figno by1}
%%%
\else
\def\ifig#1#2#3{\xdef#1{fig.~\the\figno}
\writedef{#1\leftbracket fig.\noexpand~\the\figno}%
%\figinsert\figin{\centerline{#3}}\medskip\centerline{\vbox{\baselineskip12pt
%\advance\hsize by -1truein\noindent\footnotefont{\bf Fig.~\the\figno:} #2}}
%\bigskip\endinsert
\global\advance\figno by1} \fi

%%%%%%% References %%%%%%%

\def\tilde{\widetilde}

\def\p{\partial}

\def\mod{{\rm mod}}
\def\det{{\rm det}}

%% MACROS

\def\IL{\relax{\rm I\kern-.18em L}}
\def\IH{\relax{\rm I\kern-.18em H}}
\def\IR{\relax{\rm I\kern-.18em R}}
\def\IC{\relax\hbox{$\inbar\kern-.3em{\rm C}$}}
\def\IZ{\relax\ifmmode\mathchoice
{\hbox{\cmss Z\kern-.4em Z}}{\hbox{\cmss Z\kern-.4em Z}}
{\lower.9pt\hbox{\cmsss Z\kern-.4em Z}} {\lower1.2pt\hbox{\cmsss
Z\kern-.4em Z}}\else{\cmss Z\kern-.4em Z}\fi}
\def\CM {{\cal M}}

\def\CV {{\cal V}}
\def\CO {{\cal O}}
\def\CZ {{\cal Z}}

\def\CC {{\cal C}}

\def\CS {{\cal S}}

%% MORE MACROS
\def\CM {{\cal M}}

\def\CO {{\cal O}}

\def\CV{{\cal V }}
\def\CZ {{\cal Z }}
\def\CS {{\cal S }}

\def\det{{\rm det}}
\def\Tr{{\rm Tr}}

\font\manual=manfnt \def\dbend{\lower3.5pt\hbox{\manual\char127}}

\def\IZ{\relax\ifmmode\mathchoice
{\hbox{\cmss Z\kern-.4em Z}}{\hbox{\cmss Z\kern-.4em Z}}
{\lower.9pt\hbox{\cmsss Z\kern-.4em Z}} {\lower1.2pt\hbox{\cmsss
Z\kern-.4em Z}}\else{\cmss Z\kern-.4em Z}\fi}

\def\lfm#1{\medskip\noindent\item{#1}}
\def\p{\partial}

\def\bar{\overline}
\def\CS{{\cal S}}

\def\Re{{\rm Re\, }}
\def\Im{{\rm Im\, }}

\def\rt2{\sqrt{2}}
\def\irt2{{1\over\sqrt{2}}}

\def\hat{\widehat}
%  \slashchar puts a slash through a character to represent contraction
%  with Dirac matrices. Use \not instead for negation of relations, and use
%  \hbar for hbar.
\def\slashchar#1{\setbox0=\hbox{$#1$}           % set a box for #1
   \dimen0=\wd0                                 % and get its size
   \setbox1=\hbox{/} \dimen1=\wd1               % get size of /
   \ifdim\dimen0>\dimen1                        % #1 is bigger
      \rlap{\hbox to \dimen0{\hfil/\hfil}}      % so center / in box
      #1                                        % and print #1
   \else                                        % / is bigger
      \rlap{\hbox to \dimen1{\hfil$#1$\hfil}}   % so center #1
      /                                         % and print /
   \fi}

%\DavidTX
\lref\DavidTX{ F.~David, ``Planar Diagrams, Two-Dimensional
Lattice Gravity And Surface Models,'' Nucl.\ Phys.\ B {\bf 257},
45 (1985).
%%CITATION = NUPHA,B257,45;%%
}

%\KazakovDS
\lref\KazakovDS{ V.~A.~Kazakov, ``Bilocal Regularization Of Models
Of Random Surfaces,'' Phys.\ Lett.\ B {\bf 150}, 282 (1985).
%%CITATION = PHLTA,B150,282;%%
}
%\KazakovEA
\lref\KazakovEA{ V.~A.~Kazakov, A.~A.~Migdal and I.~K.~Kostov,
``Critical Properties Of Randomly Triangulated Planar Random
Surfaces,'' Phys.\ Lett.\ B {\bf 157}, 295 (1985).
%%CITATION = PHLTA,B157,295;%%
}
%\AmbjornAZ
\lref\AmbjornAZ{ J.~Ambjorn, B.~Durhuus and J.~Frohlich,
``Diseases Of Triangulated Random Surface Models, And Possible
Cures,'' Nucl.\ Phys.\ B {\bf 257}, 433 (1985).
%%CITATION = NUPHA,B257,433;%%
}
%\DouglasVE
\lref\DouglasVE{ M.~R.~Douglas and S.~H.~Shenker, ``Strings In
Less Than One-Dimension,'' Nucl.\ Phys.\ B {\bf 335}, 635 (1990).
%%CITATION = NUPHA,B335,635;%%
}
%\GrossVS
\lref\GrossVS{ D.~J.~Gross and A.~A.~Migdal, ``Nonperturbative
Two-Dimensional Quantum Gravity,'' Phys.\ Rev.\ Lett.\  {\bf 64},
127 (1990).
%%CITATION = PRLTA,64,127;%%
}
%\BrezinRB
\lref\BrezinRB{ E.~Brezin and V.~A.~Kazakov, ``Exactly Solvable
Field Theories Of Closed Strings,'' Phys.\ Lett.\ B {\bf 236}, 144
(1990).
%%CITATION = PHLTA,B236,144;%%
}
%\DouglasDD
\lref\DouglasDD{ M.~R.~Douglas, ``Strings In Less Than
One-Dimension And The Generalized K-D-V Hierarchies,'' Phys.\
Lett.\ B {\bf 238}, 176 (1990).
%%CITATION = PHLTA,B238,176;%%
}

%\GinspargIS
\lref\GinspargIS{ P.~Ginsparg and G.~W.~Moore, ``Lectures On 2-D
Gravity And 2-D String Theory,'' arXiv:hep-th/9304011.
%%CITATION = HEP-TH 9304011;%%
}

%\MooreAG
\lref\MooreAG{ G.~W.~Moore and N.~Seiberg, ``From loops to fields
in 2-D quantum gravity,'' Int.\ J.\ Mod.\ Phys.\ A {\bf 7}, 2601
(1992).
%%CITATION = IMPAE,A7,2601;%%
}

%\KlebanovTB
\lref\KlebanovTB{ I.~R.~Klebanov and E.~Witten, ``AdS/CFT
correspondence and symmetry breaking,'' Nucl.\ Phys.\ B {\bf 556},
89 (1999) [arXiv:hep-th/9905104].
%%CITATION = HEP-TH 9905104;%%
}

%\KostovHN
\lref\KostovHN{ I.~K.~Kostov, ``Multiloop correlators for closed
strings with discrete target space,'' Phys.\ Lett.\ B {\bf 266},
42 (1991).
%%CITATION = PHLTA,B266,42;%%
}

%\DiFrancescoNW
\lref\DiFrancescoNW{ P.~Di Francesco, P.~Ginsparg and
J.~Zinn-Justin, ``2-D Gravity and random matrices,'' Phys.\ Rept.\
{\bf 254}, 1 (1995) [arXiv:hep-th/9306153].
%%CITATION = HEP-TH 9306153;%%
}

%\DornSV
\lref\DornSV{ H.~Dorn and H.~J.~Otto, ``Some conclusions for
noncritical string theory drawn from two and three point functions
in the Liouville sector,'' arXiv:hep-th/9501019.
%%CITATION = HEP-TH 9501019;%%
}
%\ZamolodchikovAA
\lref\ZamolodchikovAA{ A.~B.~Zamolodchikov and
A.~B.~Zamolodchikov, ``Structure constants and conformal bootstrap
in Liouville field theory,'' Nucl.\ Phys.\ B {\bf 477}, 577 (1996)
[arXiv:hep-th/9506136].
%%CITATION = HEP-TH 9506136;%%
}

%\TeschnerYF
\lref\TeschnerYF{ J.~Teschner, ``On the Liouville three point
function,'' Phys.\ Lett.\ B {\bf 363}, 65 (1995)
[arXiv:hep-th/9507109].
%%CITATION = HEP-TH 9507109;%%
}

%\FateevIK
\lref\FateevIK{ V.~Fateev, A.~B.~Zamolodchikov and
A.~B.~Zamolodchikov, ``Boundary Liouville field theory. I:
Boundary state and boundary  two-point function,''
arXiv:hep-th/0001012.
%%CITATION = HEP-TH 0001012;%%
}
%\TeschnerMD
\lref\TeschnerMD{ J.~Teschner, ``Remarks on Liouville theory with
boundary,'' arXiv:hep-th/0009138.
%%CITATION = HEP-TH 0009138;%%
}

%\ZamolodchikovAH
\lref\ZamolodchikovAH{ A.~B.~Zamolodchikov and
A.~B.~Zamolodchikov, ``Liouville field theory on a pseudosphere,''
arXiv:hep-th/0101152.
%%CITATION = HEP-TH 0101152;%%
}
%\PonsotNG
\lref\PonsotNG{ B.~Ponsot and J.~Teschner, ``Boundary Liouville
field theory: Boundary three point function,'' Nucl.\ Phys.\ B
{\bf 622}, 309 (2002) [arXiv:hep-th/0110244].
%%CITATION = HEP-TH 0110244;%%
}

%\SeibergNM
\lref\SeibergNM{ N.~Seiberg and D.~Shih, ``Branes, rings and
matrix models in minimal (super)string theory,''
arXiv:hep-th/0312170.
%%CITATION = HEP-TH 0312170;%%
}
%\KlebanovWG
\lref\KlebanovWG{ I.~R.~Klebanov, J.~Maldacena and N.~Seiberg,
``Unitary and complex matrix models as 1-d type 0 strings,''
arXiv:hep-th/0309168.
%%CITATION = HEP-TH 0309168;%%
}
%\KutasovFG
\lref\KutasovFG{ D.~Kutasov, K.~Okuyama, J.~Park, N.~Seiberg and
D.~Shih, ``Annulus amplitudes and ZZ branes in minimal string
theory,'' arXiv:hep-th/0406030.
%%CITATION = HEP-TH 0406030;%%
}
%\MorozovHH
\lref\MorozovHH{ A.~Morozov, ``Integrability And Matrix Models,''
Phys.\ Usp.\  {\bf 37}, 1 (1994) [arXiv:hep-th/9303139].
%%CITATION = HEP-TH 9303139;%%
}
%\GaiottoYB
\lref\GaiottoYB{ D.~Gaiotto and L.~Rastelli, ``A paradigm of
open/closed duality: Liouville D-branes and the Kontsevich
%model,''
arXiv:hep-th/0312196.
%%CITATION = HEP-TH 0312196;%%
}

%\AmbjornMY
\lref\AmbjornMY{ J.~Ambjorn, S.~Arianos, J.~A.~Gesser and
S.~Kawamoto, ``The geometry of ZZ-branes,'' arXiv:hep-th/0406108.
%%CITATION = HEP-TH 0406108;%%
}

%\BanksDF
\lref\BanksDF{ T.~Banks, M.~R.~Douglas, N.~Seiberg and
S.~H.~Shenker, ``Microscopic And Macroscopic Loops In
Nonperturbative Two-Dimensional Gravity,'' Phys.\ Lett.\ B {\bf
238}, 279 (1990).
%%CITATION = PHLTA,B238,279;%%
}

\lref\HanadaIM{ M.~Hanada, M.~Hayakawa, N.~Ishibashi, H.~Kawai,
T.~Kuroki, Y.~Matsuo and T.~Tada, ``Loops versus Matrices The
nonperturbative aspects of noncritical string,''
[arXiv:hep-th/0405076].
%%CITATION = HEP-TH 0405076;%%
}

%\DiFrancescoNW
\lref\DiFrancescoNW{ P.~Di Francesco, P.~Ginsparg and
J.~Zinn-Justin, ``2-D Gravity and random matrices,'' Phys.\ Rept.\
{\bf 254}, 1 (1995) [arXiv:hep-th/9306153].
%%CITATION = HEP-TH 9306153;%%
}

%\MooreIR
\lref\MooreIR{ G.~W.~Moore, N.~Seiberg and M.~Staudacher, ``From
loops to states in 2-D quantum gravity,'' Nucl.\ Phys.\ B {\bf
362}, 665 (1991).
%%CITATION = NUPHA,B362,665;%%
}

%\EynardSG
\lref\EynardSG{ B.~Eynard and J.~Zinn-Justin, ``Large order
behavior of 2-D gravity coupled to d < 1 matter,'' Phys.\ Lett.\ B
{\bf 302}, 396 (1993) [arXiv:hep-th/9301004].
%%CITATION = HEP-TH 9301004;%%
}

%\KazakovDU
\lref\KazakovDU{ V.~A.~Kazakov and I.~K.~Kostov, ``Instantons in
non-critical strings from the two-matrix model,''
arXiv:hep-th/0403152.
%%CITATION = HEP-TH 0403152;%%
}

\lref\Berry{ M.~V.~Berry, ``Stokes' phenomenon; smoothing a
Victorian discontinuity," Inst. Hautes \'Etudes Sci.\ Publ.\
Math.\ {\bf 68}, 211 (1988).
}

%\DijkgraafVP
\lref\DijkgraafVP{ R.~Dijkgraaf, A.~Sinkovics and M.~Temurhan,
``Universal correlators from geometry,'' arXiv:hep-th/0406247.
%%CITATION = HEP-TH 0406247;%%
}

%\AganagicQJ
\lref\AganagicQJ{ M.~Aganagic, R.~Dijkgraaf, A.~Klemm, M.~Marino
and C.~Vafa, ``Topological strings and integrable hierarchies,''
arXiv:hep-th/0312085.
%%CITATION = HEP-TH 0312085;%%
}

%\DaulBG
\lref\DaulBG{ J.~M.~Daul, V.~A.~Kazakov and I.~K.~Kostov,
``Rational theories of 2-D gravity from the two matrix model,''
Nucl.\ Phys.\ B {\bf 409}, 311 (1993) [arXiv:hep-th/9303093].
%%CITATION = HEP-TH 9303093;%%
}
%\McGreevyKB
\lref\McGreevyKB{ J.~McGreevy and H.~Verlinde, ``Strings from
tachyons: The $c = 1$ matrix reloaded,'' arXiv:hep-th/0304224.
%%CITATION = HEP-TH 0304224;%%
}

%\MartinecKA
\lref\MartinecKA{ E.~J.~Martinec, ``The annular report on
non-critical string theory,'' arXiv:hep-th/0305148.
%%CITATION = HEP-TH 0305148;%%
}

%\KlebanovKM
\lref\KlebanovKM{ I.~R.~Klebanov, J.~Maldacena and N.~Seiberg,
``D-brane decay in two-dimensional string theory,'' JHEP {\bf
0307}, 045 (2003) [arXiv:hep-th/0305159].
%%CITATION = HEP-TH 0305159;%%
}
%\McGreevyEP
\lref\McGreevyEP{ J.~McGreevy, J.~Teschner and H.~Verlinde,
``Classical and quantum D-branes in 2D string theory,''
arXiv:hep-th/0305194.
%%CITATION = HEP-TH 0305194;%%
}

%\BanksVH
\lref\BanksVH{ T.~Banks, W.~Fischler, S.~H.~Shenker and
L.~Susskind, ``M theory as a matrix model: A conjecture,'' Phys.\
Rev.\ D {\bf 55}, 5112 (1997) [arXiv:hep-th/9610043].
%%CITATION = HEP-TH 9610043;%%
}

%\FidkowskiNF
\lref\FidkowskiNF{ L.~Fidkowski, V.~Hubeny, M.~Kleban and
S.~Shenker, ``The black hole singularity in AdS/CFT,'' JHEP {\bf
0402}, 014 (2004) [arXiv:hep-th/0306170].
%%CITATION = HEP-TH 0306170;%%
}

%\MooreCN
\lref\MooreCN{ G.~W.~Moore, ``Matrix Models Of 2-D Gravity And
Isomonodromic Deformation,'' Prog.\ Theor.\ Phys.\ Suppl.\  {\bf
102}, 255 (1990).
%%CITATION = PTPSA,102,255;%%
}
%\MooreMG
\lref\MooreMG{ G.~W.~Moore, ``Geometry Of The String Equations,''
Commun.\ Math.\ Phys.\  {\bf 133}, 261 (1990).
%%CITATION = CMPHA,133,261;%%
}

%\BleherYS
\lref\BleherYS{
P.~Bleher and A.~Its,
``Double scaling limit in the random matrix model: the Riemann-Hilbert
approach,''
arXiv:math-ph/0201003.
%%CITATION = MATH-PH 0201003;%%
}

%\BrezinVD
\lref\BrezinVD{ E.~Brezin, E.~Marinari and G.~Parisi, ``A
Nonperturbative Ambiguity Free Solution Of A String Model,''
Phys.\ Lett.\ B {\bf 242}, 35 (1990).
%%CITATION = PHLTA,B242,35;%%
}

%\DavidGE
\lref\DavidGE{ F.~David, ``Loop Equations And Nonperturbative
Effects In Two-Dimensional Quantum
%Gravity,''
Mod.\ Phys.\ Lett.\ A {\bf 5}, 1019 (1990).
%%CITATION = MPLAE,A5,1019;%%
}
%\DavidSK
\lref\DavidSK{ F.~David, ``Phases Of The Large N Matrix Model And
Nonperturbative Effects In 2-D Gravity,'' Nucl.\ Phys.\ B {\bf
348}, 507 (1991).
%%CITATION = NUPHA,B348,507;%%
}

%\DouglasXV
\lref\DouglasXV{ M.~R.~Douglas, N.~Seiberg and S.~H.~Shenker,
``Flow And Instability In Quantum Gravity,'' Phys.\ Lett.\ B {\bf
244}, 381 (1990).
%%CITATION = PHLTA,B244,381;%%
}

%\DijkgraafQH
\lref\DijkgraafQH{ R.~Dijkgraaf, ``Intersection theory, integrable
hierarchies and topological field theory,'' arXiv:hep-th/9201003.
%%CITATION = HEP-TH 9201003;%%
}

%\KostovXI
\lref\KostovXI{ I.~K.~Kostov, ``Conformal field theory techniques
in random matrix models,'' arXiv:hep-th/9907060.
%%CITATION = HEP-TH 9907060;%%
}

\lref\segalwilson{G. Segal and G. Wilson, ``Loop groups and equations of
KdV type,'' Publ. IHES {\bf 61}(1985) 1. }

%\writedefs

%\draftmode

\newbox\tmpbox\setbox\tmpbox\hbox{\abstractfont }
\Title{\vbox{\baselineskip12pt\hbox to\wd\tmpbox{\hss }}\hbox{
PUPT-2129 }}
{\vbox{\centerline{Exact vs. Semiclassical Target
Space}\smallskip \centerline{ of the Minimal String }}}
\smallskip
\centerline{  Juan Maldacena,$^1$ Gregory Moore,$^2$ Nathan
Seiberg$^1$ and David Shih$^3$}
\smallskip
\bigskip
\centerline{$^1${\it School of Natural Sciences, Institute for
Advanced Study, Princeton, NJ 08540, USA}}
\medskip
\centerline{$^2${\it Department of Physics, Rutgers University,
Piscataway, NJ 08854, USA}}
\medskip
\centerline{$^3${\it Department of Physics, Princeton University,
Princeton, NJ 08544, USA}}
\bigskip
\vskip 1cm

\noindent We study both the classical and the quantum target space
of $(p,q)$ minimal string theory, using the FZZT brane as a probe.
By thinking of the target space as the moduli space of FZZT
branes, parametrized by the boundary cosmological constant $x$, we
see that classically it consists of a Riemann surface $\CM_{p,q}$
which is a $p$-sheeted cover of the complex $x$ plane. However, we
show using the dual matrix model that the exact quantum FZZT
observables exhibit Stokes' phenomenon and are entire functions of
$x$. Along the way we clarify some points about the semiclassical
limit of D-brane correlation functions. The upshot is that
nonperturbative effects modify the target space drastically,
changing it from $\CM_{p,q}$ to the complex $x$ plane. To
illustrate these ideas, we study in detail the example of
$(p,q)=(2,1)$, which is dual to the Gaussian matrix model. Here we
learn that the other sheets of the classical Riemann surface
describe instantons in the effective theory on the brane. Finally,
we discuss possible applications to black holes and the
topological string.

\vskip .5cm \Date{August, 2004}

\newsec{Introduction}

Minimal string theories, or $(p,q)$ minimal CFTs coupled to
Liouville theory, are important examples of tractable, exactly
solvable models of quantum gravity. These models are interesting
laboratories for the study of string theory because, despite their
simplicity, they contain many of the features of critical string
theory, including D-branes, holography and open/closed duality.
First solved using the dual matrix model description
\refs{\DavidTX\KazakovDS\KazakovEA\AmbjornAZ\DouglasVE\GrossVS
\BrezinRB-\DouglasDD} (for reviews, see e.g.
\refs{\GinspargIS,\DiFrancescoNW}), recent progress in the study
of Liouville theory
\refs{\DornSV\ZamolodchikovAA\TeschnerYF\FateevIK\TeschnerMD
\ZamolodchikovAH-\PonsotNG}
has led to a greatly improved understanding of minimal string
theory from the worldsheet perspective
\refs{\McGreevyKB\MartinecKA\KlebanovKM\McGreevyEP\KlebanovWG
\SeibergNM\GaiottoYB\HanadaIM\KutasovFG-\AmbjornMY}.

One limitation of minimal string theory, however, has so far been
the lack of a well-developed target space interpretation. In this
paper, we will take the first steps towards a solution of this
problem. Naively, the target space of minimal string theory is
just the worldsheet Liouville field $\phi$. However, it is common
in string theory that different target spaces can have the same
physics.  In our case an equivalent description involves the free
scalar field $\tilde\phi$ related to $\phi$ by the non-local
B\"acklund transformation (similar to T-duality). An important
question which we will address is the distinction between the
classical target space and the nonperturbative, quantum target
space.  We will see that they are quite different.

Our point of view (which was used among other places in \BanksVH)
is that a better effective description of target space can emerge
out of the moduli space of D-branes.  The advantage of this point
of view is that it can capture all the nonperturbative corrections
to the target space. We should point out, however, that different
branes can lead to different target spaces.  For example in
compactification on a circle D0-branes probe the circle, while the
D1-branes probe the dual circle.

Minimal string theories have D-branes (the FZZT branes) labelled
by a continuous real parameter
\eqn\defx{
x = \mu_B
}
(the boundary cosmological constant). We wish to interpret $x$ as
a target space coordinate, and the picture we have in mind is as
follows.
%\foot{We start with the conformal backgrounds described in
%the introduction. Below we will discuss more general backgrounds
%as deformations of this one.}
The minisuperspace wavefunction of the FZZT brane suggests that it
is a D-brane in $\phi$ space stretching from $\phi=-\infty$ and
dissolving at $\phi\sim -{1\over b}\log x$ (where $b=\sqrt{p\over
q}$ is the Liouville coupling constant). Therefore, the tip of the
FZZT brane at $\phi\sim -{1\over b}\log x$ acts as a point-like
probe of the Liouville direction.\foot{This intuition can be made
more precise using the equivalent description in terms of the
B\"acklund field $\tilde \phi$. Here the FZZT brane corresponds to
a {\it Dirichlet} boundary condition on $\tilde \phi$ \SeibergNM.}
It has the virtue of being able to penetrate into the
strong-coupling region $\phi\to+\infty$, where one might expect
there to be significant modifications to the classical target
space. So in this description, target space is parametrized by the
coordinate $x$. Large positive $x$ corresponds to the
weak-coupling region of the Liouville direction, while $x$ of
order one corresponds to strong coupling.

To see how the worldsheet dynamics of the Liouville field modifies
the naive target space, it is useful to analytically continue $x$
to complex values.  In the semiclassical approximation the
D-branes are not single valued as a function of $x$ and are
labelled by a point in a finite multiple cover of the $x$-plane.
This multiple cover corresponds to a Riemann surface $\CM$ which
can be described as follows.  In terms of the disk amplitude $
\Phi$ we define
 \eqn\defy{y=\partial_x \Phi}
or equivalently
 \eqn\calZin{\Phi=\int^x y(x') dx'}
Then, $x$ and $y$ satisfy an algebraic equation
 \eqn\alge{F(x,y)=0}
which describes the Riemann surface $\CM$ in ${\Bbb C}^2$.  The
parameters of the polynomial $F(x,y)$ which determine the complex
structure of $\CM$ depend on the parameters of the minimal string.
For large $x$ and $y$ the equation $F(x,y)=0$ becomes of the form
$x^q \approx y^p$ for integer $p$ and $q$.  We will refer to the
corresponding Riemann surface as $\CM_{p,q}$.

Physically, what is happening is that we start probing the target
space at large positive $x$, where the Liouville field is weakly
coupled and the classical target space is a good description.
Next, we bring the FZZT brane probe into the strong-coupling
region $x$ of order one, and we find a branch point at $x=-1$.
This branch point is a sign that the target space is modified due
to strong-coupling effects on the worldsheet. It suggests that we
analytically continue into the complex $x$ plane, through the
branch cut ending at $x=-1$, and into the other sheets of the
Riemann surface. Correspondingly, the moduli space of FZZT branes
must be enlarged from the complex $x$ plane to the Riemann surface
$\CM_{p,q}$. In this way, we obtain $\CM_{p,q}$ for the
semiclassical target space of the minimal string.\foot{It may seem
strange that we started with a one dimensional target space
consisting of $\phi$, and we ended up with a two dimensional
target space consisting of $\CM_{p,q}$. However, the fact that all
physical quantities depend holomorphically on $\CM_{p,q}$ suggests
that there is still a sense in which the target space is one
dimensional.}

A special situation occurs when $\CM_{p,q}$ has genus zero. This
happens for backgrounds without ZZ branes
\refs{\SeibergNM,\KutasovFG}. Since here $\CM_{p,q}$ has genus
zero, it can be uniformized by a single complex parameter $z$;
i.e. there is a one to one map between the complex $z$ plane and
$\CM_{p,q}$.  Then we can express the values of $x$ and $y$ as
polynomials of degrees $p$ and $q$ in $z$
 \eqn\xyzp{x=x_p(z)\ , \qquad y=y_q(z)}
The parameters in these polynomials depend on the closed string
background which is labelled by the coefficients of closed string
operators $t=(t_1,t_2,\dots)$.  Of particular importance among
these parameters are the coefficient of the lowest dimension
operator
\eqn\lowestcplingin{
\tau \equiv t_1
}
and the worldsheet cosmological constant $\mu$ (in unitary
worldsheet theories $\tau =\mu$, but in general they are
different). The uniformizing parameter $z$ has the worldsheet
interpretation as the one point function on the disk of the lowest
dimension operator
 \eqn\zexp{z= \partial_\tau \Phi=
 \partial_\tau\left(\int^x y dx\right)\Big|_{x}}
Below we will relate this expression to various points of view of
the minimal string, in particular the connection to integrable
hierarchies.

Most of the discussion in this paper will concern the class of
backgrounds where $\CM_{p,q}$ has genus zero.  We expect our main
results to apply to the more generic backgrounds with arbitrary
Riemann surface and we will comment on this below.

A further useful specialization of the background is to the {\it
conformal backgrounds } of \MooreIR. These are backgrounds without
ZZ branes in which all the closed string couplings $t$, with the
exception of the worldsheet cosmological constant $\mu$, have been
set to zero.
%The worldsheet description of these theories is in
%terms of a minimal conformal field theory coupled to Liouville
%theory.
In this class of theories explicit worldsheet
calculations can be performed, leading to checks of the results in
more general backgrounds. Here the more general expressions
\alge,\xyzp\ become \SeibergNM\ (see also \DaulBG)
 \eqn\Mpq{\eqalign{
 &F(x,y)=T_p(y/C)-T_q(x) =0\cr
 }}
and
\eqn\xyconfintro{\eqalign{
 &x=T_p(z),\qquad
 y=C T_q(z) \cr}}
Here $C$ is a normalization factor which we will determine below,
$T_n(\cos \theta) = \cos(n\theta)$ is a Chebyshev polynomial of
the first kind, and for simplicity we have set $\mu=1$.

It is natural to expect that this picture of the classical target
space is modified only slightly when perturbative effects in the
string coupling $g_s=\hbar$ are taken into account. As we will
see, however, nonperturbative effects have important consequences.
To study the quantum target space, we turn in the first part of
section 2 to the nonperturbative description of FZZT branes
afforded by the dual matrix model. In the matrix model, FZZT
branes are described by insertions of the exponentiated
macroscopic loop operator
\refs{\BanksDF\MooreIR\MooreAG-\KostovHN}
 \eqn\fzztmat{\Psi(x)\sim e^{\Tr\log(x-M)}= \det(x-M) }
into the matrix integral.  We use the results of \MorozovHH\ to
compute the correlator of any number of FZZT branes. Taking the
continuum limit, we show that the correlators become
\eqn\fzztcorrintro{ \left\langle
 \prod_{i=1}^{n}\Psi(x_i)\right\rangle
 = {\Delta(d_j)\over \Delta(x)}\prod_{i=1}^{n}\psi(x_i,t)
 }
where \eqn\psidefintro{ \psi(x,t)=\langle \Psi(x)\rangle } is the
FZZT partition function, which depends on the closed-string
couplings $t=t_1,t_2,\dots$; $\Delta$ denotes the Vandermonde
determinant; and $d_j$ is shorthand for the action of
$d=\hbar\partial_\tau$ on $\psi(x_j,t)$. (Note that it does {\it
not} refer to differentiation with respect to $x_j$.) We argue
that the denominator can be removed by thinking of the FZZT branes
as wavefunctions (half-densities), with the result that the branes
become fermionic.

In the classical $\hbar\to 0$ limit, we show that the FZZT
correlators reduce to
\eqn\corrclintro{
 \lim_{\hbar\to 0}\left\langle\prod_{i=1}^{n}\Psi(x_i)\right\rangle
 = {\Delta(z)\over \Delta(x)}\prod_{i=1}^{n}\Psi_{cl}(z)
 }
where
 \eqn\Psiclintro{ \Psi_{cl}(z) =
 x'(z)^{-1/2}e^{\int^{x(z)}y(x)dx/\hbar} }
is the semiclassical approximation to the FZZT partition function.
We also provide a worldsheet interpretation for the various
factors in \corrclintro\ -- the first factor ${\Delta(z)\over
\Delta(x)}$ comes from annulus diagrams between different FZZT
branes, while the semiclassical wavefunctions $\Psi_{cl}(x_i)$
come from the disk amplitude and the annulus between the same
brane.  More generally, we interpret these expressions as a change
in the measure of the D-branes $\Psi(x)$ to a fermion on the
Riemann surface $\CM_{p,q}$.

Using the fact that $\psi(x,t)$ is a Baker-Akhiezer function of
the KP hierarchy, which is actually an {\it entire} function of
$x$ \refs{\MooreCN,\MooreMG}, it follows that the exact FZZT
correlators \fzztcorrintro\ are all entire functions of $x$. This
is in spite of the fact that the classical correlators
\corrclintro\ are clearly functions on $\CM_{p,q}$. Evidently, the
quantum target space differs significantly from the classical
target space. Whereas the latter comprised the Riemann surface
$\CM_{p,q}$, the former consists of only the complex $x$ plane!

In section 2.4, we analyze additional FZZT observables (the
quantum resolvents) and show that they are also entire functions
of $x$. Finally, we rederive the WKB approximation \Psiclintro\
using the fact that $\psi(x,t)$ is a Baker-Akhiezer function. We
note that the asymptotics exhibit ``level crossing'' behavior at
large negative $x$.
%%%
Here, by level crossing we mean simply that there is a branch cut
along the negative real axis with different values of $\psi(x,t)$
above and below the cut. Below, when this approximation of
$\psi(x,t)$ will be associated with saddle points in an integral,
we will see that two saddle points exchange dominance there.

In section 3, we illustrate our general arguments with the
simplest example of minimal string theory, namely the topological
$(p,q)=(2,1)$ model. This is dual to the Gaussian matrix model,
and we show that the FZZT partition function is expressed in terms
of the Airy function. By representing the insertion of a D-brane
as a Grassmann integral in the matrix model,
 we give a direct and simple proof of the
equivalence between the $n\times n$ Kontsevich model and the
double-scaled $(2,1)$ model with $n$ FZZT branes. That is, we show
that in the continuum double scaling limit
 \eqn\equivintro{ \left\langle
 \prod_{i=1}^{n}\Psi(x_i)\right\rangle \to \int dS\
 e^{\Tr\left(i  S^3/3+ i\hbar^{-2/3} (X+\tau)S\right)} }
with $S$ and $X$ $n\times n$ Hermitian matrices and $\tau$, which
can be absorbed in $X$, is the coupling constant of the theory.
The eigenvalues of $X$ are $x_1,\dots,x_n$ after an appropriate
shift and rescaling in the double scaling limit (see below). Using
this approach, we see very directly how the matrix $S$ of the
$n\times n$ Kontsevich model is the effective degree of freedom
describing open strings stretched between $n$ FZZT branes
\GaiottoYB.

We continue our study of the $(2,1)$ model in section 4, focusing
now on the effective theory on the FZZT brane, which is described
by the Airy integral \equivintro\ with $n=1$. An analysis of this
integral using the stationary phase method reveals several new
facts. To begin, we show how the other sheets of the classical
moduli space can be viewed as saddle points in the integral
describing the FZZT partition function. Therefore, they can be
thought of as instantons in the effective theory on the brane. We
expect this conclusion to hold for all values of $(p,q)$.

A more careful stationary phase analysis of the Airy integral
illustrates the general mechanism by which the target space is
modified nonperturbatively. Exponentially small quantities --
neglected in perturbative string theory -- can become large upon
analytic continuation, and these large corrections ``erase'' the
branch cuts and monodromies of the Riemann surface in the exact
answer. The essence of the replacement of the semiclassical target
space $\CM_{p,q}$ by the humble complex $x$-plane is thus what is
known as Stokes' phenomenon. Generally speaking, Stokes'
phenomenon  is the fundamental fact that the analytic continuation
of an asymptotic expansion can differ from the asymptotic
expansion of an analytic continuation. In our case, the Riemann
surface is extracted by working in the classical approximation to
string theory (thus taking the leading term in an asymptotic
expansion in $g_s=\hbar$) and then considering the analytic
continuation. Thus, the Riemann surface arises from the analytic
continuation of the asymptotics. Thanks to the matrix model we can
study the analytic continuation of the exact nonperturbative
answers for amplitudes directly.  The fact that the FZZT
amplitudes are entire shows that the Riemann surface
``disappears'' nonperturbatively.

In terms of the saddle point analysis of the field theory living
on the FZZT branes, Stokes' phenomenon is exhibited in two ways.
The first, more trivial way, occurs when the parameter $x$ is
varied across what is known as an {\it anti-Stokes' line}. This
can be thought of as a first-order phase transition where two
contributing saddle points exchange dominance. The story is
incomplete, however, if we simply consider only the anti-Stokes'
lines. In addition, there is also a more subtle phenomenon
happening along what are called {\it Stokes' lines}. As $x$ is
varied across a Stokes' line, a subdominant saddle abruptly ceases
to contribute to the exact answer. This phenomenon is most
dramatic when we continue to vary $x$ and the missing saddle
becomes the {\it dominant} saddle, even though it is still not
contributing to the integral. In terms of the path integral
describing the effective theory on the brane, what is happening is
that one simply cannot deform the contour of integration to pass
through that saddle. We will discuss this in more detail in
Appendix B.

In section 5, we comment on the issues involved in generalizing to
other backgrounds. Among other things, we show using the FZZT
partition function that not all values of $(p,q)$ correspond to
nonperturbatively consistent backgrounds with a double-scaled
matrix model that is bounded from below. We deduce a bound
\eqn\boundintro{
\sin{\pi q\over p} > 0
}
that must be satisfied in order for the corresponding background
to exist. For instance, when $p=2$ only the $(p,q)=(2,2m-1)$
models with $m$ odd exist nonperturbatively, while the models
based on the unitary discrete series with $q=p+1$ never exist
nonperturbatively.

Finally, section 6 contains a possible analogy with the work of
\FidkowskiNF\ on the physics behind black hole horizons and
possible implications for the topological string approach of
\refs{\AganagicQJ}. In appendix A, we review the Lax formalism of
minimal string theory (the operators $P$ and $Q$ and the string
equation $[P,Q]=\hbar$), and we present new results concerning the
geometrical interpretation of its classical limit. Appendix B, as
mentioned above, contains a brief review of Stokes' phenomenon
along the lines of \Berry, while appendix C contains the results
of a numerical analysis of $(p,q)=(2,5)$.

\newsec{The Quantum Target Space: FZZT Branes in the Matrix Model}

\subsec{FZZT correlators at finite $N$}

As is well-known, $(p,q)$ minimal string theory possesses a dual
matrix model description. For $p=2$, the dual matrix model
consists of an $N\times N$ Hermitian matrix $M$ with potential
$V(M)$ and coupling $g$,
 \eqn\mmdef{\CZ(g) = \int dM\ e^{-{1 \over g} \Tr V(M)}}
while for $p>2$ one needs to use an analogously defined two-matrix
model (for recent discussion of the two-matrix model and
references, see \KazakovDU):
\eqn\mmdefII{
 \tilde \CZ(g) = \int dM d\tilde M\ e^{-{1\over g}\left(\Tr V(M)+\Tr W(\tilde
 M)-{\Tr} M\tilde M\right)}
 }
Here the measures $dM$ and $d\tilde M$ include a factor of the
volume of $U(N)$.

In the matrix model, macroscopic loops are created by insertions
of the operator
 \eqn\mmloop{ W(x) = {1\over N}\Tr\log(x-M)}
in the matrix integral.\foot{In the two-matrix model, there is
another loop made out of $\tilde M$. It corresponds to the ``dual"
FZZT brane and classically is related to the loop \mmloop\ by a
Legendre-type transform \KazakovDU. We will discuss the
interpretation of this dual loop in section 6.} For instance, the
large $N$ limit of $\langle W(x)\rangle$ corresponds to the FZZT
disk amplitude $\Phi$ (up to a polynomial in $x$), and the matrix
model resolvent
\eqn\mmresolvent{ R(x) =
\partial_x \langle W(x)\rangle = {1\over N}\left\langle \Tr
{1\over x-M}\right\rangle } corresponds to $y(x)$ (again up to a
polynomial in $x$).

The full FZZT brane obviously does not correspond to a single
macroscopic loop in the worldsheet. Rather, we must include
contributions from worldsheets with any number of boundaries. This
is accomplished by exponentiating $W(x)$, whereby the full,
nonperturbative FZZT brane is represented by a determinant
operator \eqn\mmdet{ e^{N W(x)} = \det(x-M) } in the matrix model.
We can also write this determinant as a Grassmann integral over
$N$ complex fermions $\chi_i$
\eqn\detrew{ \det(x-M) = \int d\chi d\chi^\dagger \
e^{\chi^\dagger(x-M)\chi} } In
\refs{\McGreevyKB\MartinecKA\KlebanovKM-\McGreevyEP,\KutasovFG},
the matrix $M$ of the one-matrix model was interpreted as
describing the (bosonic) open strings stretched between the $N$
condensed ZZ branes in the Fermi sea. Meanwhile, the $\chi_i$ are
taken to represent {\it fermionic} open strings stretched between
the FZZT brane and the $N$ ZZ branes \KutasovFG.

%It is important to emphasize that physical observables involving
%FZZT branes should be calculated using the determinant operator
%\mmdet\ rather than the macroscopic loop operator \mmloop. A good
%example of this rule is the
Now consider the correlation function of any number of FZZT
branes, which nonperturbatively is given by a product of
determinants. Amazingly, this can be explicitly evaluated in both
the one and two matrix models \MorozovHH. The answer is
\eqn\fzztcorrans{
\left\langle
\prod_{i=1}^{n}\det(x_i-M)\right\rangle
 = {\det(
P_{N+i-1}(x_j) )\over \Delta(x)}
}
Here $\Delta(x)=\prod_{i<j}(x_i-x_j)$ is the Vandermonde
determinant, $P_k(x)$ are the orthogonal polynomials of the matrix
model (or bi-orthogonal polynomials associated to $M$ in the
two-matrix model) with leading coefficient $1$, and the indices
$i$ and $j$ in \fzztcorrans\  run between $1$ and $n$. The
simplest case of the general formula \fzztcorrans\ is the FZZT
partition function. This is given by a single orthogonal
polynomial:
\eqn\fzztpart{
\langle \det(x-M)\rangle = P_N(x)
}
Below, we will take the continuum limits of \fzztcorrans\ and
\fzztpart, and we will see how the perturbative loop correlators
can be recovered.

Before we proceed, let us briefly mention an interpretation of the
FZZT correlator \fzztcorrans\ that will be useful in the next
subsection. First, we need to write the LHS of \fzztcorrans\ more
compactly in the following way
\eqn\fzztcorrrew{
\left\langle \prod_{i=1}^{n}\det(x_i-M) \right\rangle =
\left\langle \det(X\otimes I_N - I_n\otimes M)\right\rangle
}
where $I_N$ and $I_n$ denote the $N\times N$ and $n\times n$
identity matrices respectively, and $X$ is understood to be an
$n\times n$ Hermitian matrix with eigenvalues $x_1,\dots,x_n$. Now
notice that if we square $\det(X\otimes I_N - I_n\otimes M)$,
multiply by $e^{-{1\over g}(\Tr V(M)+\Tr V(X))}$, and integrate
over $X$ and $M$, we obtain the $(N+n)\times (N+n)$ matrix
integral with no insertions of FZZT branes, i.e.
 \eqn\fzztcorrsq{
 \int dX dM\ e^{-{1\over g}(\Tr V(M)+\Tr V(X))} \det(X\otimes I_N -
 I_n\otimes M)^2 = \int d\hat M\ e^{-{1\over  g} \Tr V(\hat M)}
 }
where $\hat M$ is an $(N+n)\times (N+n)$ Hermitian matrix. (As
before, the integration measures include factors of the volume of
the relevant unitary group.) The meaning of \fzztcorrsq\ is that
the FZZT creation operator $\det(X\otimes I_N-I_n\otimes M)$ acts
as a kind of wavefunction (half-density) on the space of Hermitian
matrices $X$.

The motivation for interpreting $\det(X\otimes I_N-I_n\otimes M)$
as a wavefunction in $X$-space is that it allows us to think of
the FZZT branes as fermions. To see this, recall that the measure
for an integral in $X$ space is
 \eqn\measure{ dX = dU \prod_{i=1}^{n}dx_i\ \Delta(x)^2 }
where $U$ is an $n\times n$ unitary matrix (and the measure is
such that $\int dU = 1$). Hence, a half-density $(dX)^{1/2}$
carries with it a factor of $\Delta(x)$, which is precisely what
is needed to cancel the denominator of \fzztcorrans.\foot{The
annulus diagram is the logarithm of the $n=2$ version of
\fzztcorrans. In \KutasovFG, where this diagram was calculated, it
was pointed out that the term associated with the denominator of
\fzztcorrans\ is independent of the coupling constants and
therefore could be removed. Here we see a more geometric way of
deriving this fact.} (Put differently, the factor $\Delta(x)$
plays a role analogous to that of cocycles in vertex operator
algebra theory, by enforcing the correct statistics of the
determinant operator.) This leaves the numerator of \fzztcorrans,
which is obviously antisymmetric under interchange of the $x_i$'s.
Therefore, the FZZT branes become fermionic.

\subsec{FZZT correlators in the continuum limit}

Now let us take the large $N$ double-scaling limit of
\fzztcorrans\ to obtain the D-brane correlators of minimal string
theory. For simplicity, we start with the $n=1$ case \fzztpart. As
in \fzztcorrsq, to have a well-defined scaling limit we must
consider not the determinant, but rather the following operator
\refs{\MooreCN,\GinspargIS}
 \eqn\renorm{ \Psi(x) = {1\over \sqrt{h_N}} e^{- V(x)/2g}\det(x-M)}
where $V(x)$ is the matrix model potential, and $h_{N}$ is a
normalizing constant. (Some rigorous results on the double-scaled
limit of the orthonormal wavefunctions have been derived in
\BleherYS.) This converts the orthogonal polynomials in
\fzztcorrans\ to orthonormal wavefunctions with measure $dx$. Then
the FZZT partition function in the double-scaling limit is given
by a function of $x$ and the background closed-string couplings
$t=(t_1,t_2,\dots)$,
 \eqn\psirenorm{ \langle \Psi(x)\rangle = \psi(x,t) }
which is characterized by the requirement that it satisfy the
differential equations
\eqn\QPonfzzt{ Q\psi(x,t) = x\psi(x,t),\qquad P\psi(x,t) =
\hbar\partial_x\psi(x,t)
}
with $P\propto d^q+\dots$ and $Q\propto d^p+\dots$ differential
operators in $d=\hbar\partial_\tau$.\foot{It is common in the
literature to denote the lowest dimension coupling by $x$. We
denote it here by $\tau$.} (Note that the derivative $d$ is taken
at fixed values of $x, t_{j>1}$.) $P$ and $Q$ are known as Lax
operators, and they are determined by the string equation
\eqn\streqn{ [P,Q]=\hbar}
In appendix A, we review the properties of $P$ and $Q$ and present
new results concerning the geometric interpretation of these
operators in the classical limit. For a more pedagogical
introduction to the Lax formalism and integrable hierarchies, see
e.g.\ \refs{\DiFrancescoNW, \DijkgraafQH}.

Note that the differential equations \QPonfzzt\ do not specify
$\psi(x,t)$ uniquely. In nonperturbatively consistent models, we
will see below that this ambiguity can be completely fixed, in
part by the boundary condition that $\psi(x,t)$ be real and
exponentially decreasing as $x\to +\infty$.

In the literature on integrable systems, the function $\psi(x,t)$
is referred to as the ``Baker-Akhiezer function" of the associated
KP hierarchy defined by $P$ and $Q$. Here we see that it has a
simple, physical interpretation in minimal string theory as the
FZZT partition function. For our present purposes, the most
important property of the Baker-Akhiezer function is the
non-trivial fact that it (along with all of its derivatives) is an
{\it entire function} of $x$.\foot{This fact was  proven in
\refs{\MooreCN,\MooreMG} by writing the string equation in an
equivalent form as an equation for a {\it flat} holomorphic vector
bundle on the space of $x, t_k$. The connection on this vector
bundle is polynomial in $x$. The Baker-Akhiezer function is used
to make a covariantly constant frame. From the equation $({d\over dx} -
A_x)\tilde \Psi=0$, where $\tilde \Psi$ is the frame, it follows,
via the path-ordered exponential, that $\tilde \Psi $ is entire in
$x$.} We will see momentarily that this has dramatic consequences
for the quantum moduli space of FZZT branes.

In the double-scaling limit, it is not difficult to show that the
general FZZT correlator \fzztcorrans\ becomes
 \eqn\fzztcorrdbl{
 \left\langle \prod_{i=1}^{n}\Psi(x_i) \right\rangle =
 {\Delta(d_j)\over \Delta( x )}\prod_{i=1}^n \psi(x_i,t) } where
the notation $d_j$ is shorthand for the action of $d=\hbar \p_\tau$ on  $\psi(x_j,t)$.
 To derive
\fzztcorrdbl\ use the fact that the increase of index on $P_k$
becomes a derivative with respect to $\tau$ to leading order in
the double-scaling parameter $\epsilon$. Then as $\epsilon\to0$,
only the Vandermonde determinant of derivatives with respect to
the index survives. We conclude from \fzztcorrdbl\ that the
correlator of any number of FZZT branes reduces, in the continuum
limit, to a product of Baker-Akhiezer functions $\psi(x_i,t)$ and
their derivatives.

\subsec{Comparison with the semiclassical limit}

Having obtained the exact D-brane correlators \fzztcorrdbl, it is
straightforward to take their semiclassical limit and show how the
perturbative loop amplitudes can be recovered. For this, we will
need a result from subsection 2.5, namely that as $\hbar\to 0$,
the Baker-Akhiezer function becomes an eigenfunction of $d$ with
eigenvalue $z$. (As discussed in the introduction, the global
uniformizing parameter $z$ exists only in the backgrounds without
ZZ branes.) Therefore, the semiclassical limit of \fzztcorrdbl\ is
simply
 \eqn\fzztcorrclass{
 \lim_{\hbar\to 0}\left\langle \prod_{i=1}^{n}\Psi(x_i)
 \right\rangle = {\Delta(z)\over \Delta( x )}\prod_{i=1}^n
  \Psi_{cl  }(z_i)
 }
The worldsheet description of the various terms appearing in
\fzztcorrclass\ is as follows. Recall that we can think of the
FZZT creation operator $\Psi(x) \sim e^{W(x)/\hbar}$ as the
exponentiated macroscopic loop operator. Then the first factor
${\Delta(z)\over \Delta( x )}$ is the exponentiated contribution
of the annulus diagrams with the ends of the annulus ending on
different branes. This is consistent with the explicit worldsheet
calculation in conformal backgrounds which leads to the connected
annulus amplitude \KutasovFG
\eqn\annulus{
\langle W(x)W(x')\rangle_{\rm c,\, ann} =\log\left({z-z'\over
x-x'}\right)
}
The other diagrams that contribute at this order in $\hbar\,$ are
the disk diagram \calZin\ and the annulus diagram with the two
ends on the same brane\foot{The factor of a half comes from the
fact that the open strings are ending on the same brane.}
\eqn\annulusII{
\lim_{x'\to x}{ 1 \over 2} \langle W(x)W(x')\rangle_{\rm c,\, ann}
=\lim_{z'\to z} {1 \over 2} \log\left({z-z'\over
x(z)-x(z')}\right)=-{1\over 2} \log \partial_z x(z)
}
These diagrams combine to give the WKB wave functions in
\fzztcorrclass:
 \eqn\clasw{\eqalign{
 \Psi_{cl}(z)&= f(z) e^{\Phi(z) / \hbar}\cr
 \Phi(z) &= \int ^{x(z)} y dx \cr
 f(z)& = {1\over\sqrt{ \partial_z x(z)}}}}
As is common in WKB wavefunctions, the prefactor $f(z)$ is a one
loop correction. In our case it arises from an open string loop
which is the annulus diagram.  Finally, note that higher genus
diagrams are suppressed in the $\hbar \to 0$ limit in
\fzztcorrclass.\foot{We would like to stress that the leading
order expressions \fzztcorrclass, \clasw\ are correct in any
background without ZZ branes and not only in the conformal
backgrounds.  The only fact that is needed is that the Riemann
surface $\CM_{p,q}$ can be uniformized by the complex parameter
$z$; i.e.\ that it has genus zero.  We will return to this WKB
wavefunction in section 2.5.}

The difference between the classical result \fzztcorrclass\ and
the exact quantum result \fzztcorrdbl\ is at the heart of our
analysis. The classical answer is obviously defined on a multiple
cover of the complex $x$ plane, since for the same $x$, there can
be $p$ different values of $z$. On the other hand, since
$\psi(x,t)$ and its $\tau$ derivatives are entire functions of
$x$, the exact expressions \fzztcorrdbl\ for the FZZT correlators
are actually entire in the complex $x$ plane. In other words,
there are no branch cuts or other singularities that necessitate
analytic continuation to other sheets. Apparently, the
semiclassical target space $\CM_{p,q}$ disappears when one takes
nonperturbative effects into account!

In the next subsection, we will check our picture of the quantum
target space by computing the quantum resolvent and showing that
it is also an entire function of $x$. However, we would like to
first mention another perspective on the semiclassical correlator
\fzztcorrclass\ and how this is modified in the exact answer.
Recall that the FZZT brane could be thought of as a half density
multiplied by $(dx)^{1/2}$. Thus the annulus factor $f(z)$ can be
interpreted as a measure factor implementing a transformation from
$x$ to $z$. By the same token, we can think of the correlator of
$n$ FZZT branes as a half-density multiplied by $(dX)^{1/2}$ where
$X$ is an $n\times n$ matrix with eigenvalues $x_i$ (see
\measure). Then the transformation of this half-density to
$Z$-space, with $Z$ an $n\times n$ matrix whose eigenvalues are
$z_i$, must include a factor of the Jacobian
\eqn\Jac{
\left|{\partial Z\over\partial X}\right|^{1/2} ={\Delta(z)\over
\Delta (x)}\prod_{i=1}^{n}f(z_i)
}
But according to the discussion above, this is precisely the
contribution of the annulus to the correlator! Thus we have shown
that
\eqn\reexpress{
\lim_{\hbar\to 0}\left\langle \prod_{i=1}^{n}\Psi(x_i)
 \right\rangle (dX)^{1/2} = e^{\Tr\, \Phi(Z)} (dZ)^{1/2}
}
with $\Phi(z)$ as in \clasw. Evidently, the classical correlators
reduce to extremely simple expressions in $z$-space.

These simple expressions suggest there should be an equally simple
formalism underlying the classical theory. One possibility was
alluded to above, namely that instead of thinking of these
correlators as half densities we can equivalently think of them as
fermions. Then \reexpress\ indicates that {\it in the
semi-classical limit}, there is a sense in which the FZZT branes
are actually fermions on the Riemann surface $\CM_{p,q}$.
(Remember that the $z$ plane covers $\CM_{p,q}$ exactly once.)
Such fermions are common in the matrix model literature (for
reviews, see e.g.\ \refs{\GinspargIS,\MorozovHH,\DijkgraafQH,
\KostovXI}). However, our general discussion suggests that this
simple picture cannot be correct in the full nonperturbative
theory, in which the Riemann surface $\CM_{p,q}$ is replaced by
the complex $x$ plane. We return to this point at the end of
section 6.

\subsec{The analytic structure of the quantum resolvent}

Although we have shown that physical observables involving the
determinant operator are entire functions of $x$, it remains to be
seen whether the same is true for the resolvent $R(x)$ defined in
\mmresolvent. To all orders in $\hbar$ the resolvent is expected
to exhibit monodromy and have various branch cuts in the complex
$x$ plane. But in the dual string theory, the resolvent (or rather
its integral) corresponds to the vacuum amplitude of a worldsheet
with one boundary and an arbitrary number of handles. Thus we
might expect that nonperturbative effects drastically modify the
classical resolvent, just as they modified the classical
determinant correlators \fzztcorrclass.

For simplicity, we will limit ourselves to the one-matrix model,
which describes the theories with $p=2$. Then the matrix integral
defining the resolvent can be easily reduced, using the
determinant formula \fzztcorrans, to a single integral in terms of
orthogonal polynomials:
\eqn\resred{ R(x) = \int_{-\infty}^{\infty} {\rho_N(\lambda)\over
x-\lambda} } where
\eqn\rhodef{ \rho_N(\lambda) ={1\over N}\sqrt{h_N\over h_{N-1}}  \Big(
\psi_{N-1}(\lambda)\psi_{N}'(\lambda)-\psi_{N-1}'(\lambda)\psi_{N}(\lambda)\Big)
}
with $\psi_k(\lambda) = {1\over \sqrt{h_k}} e^{-
V(\lambda)/2g}P_k(\lambda)$ the orthonormal wavefunctions of the
matrix model and $\psi' = {d\over d\lambda}\psi$.
As we discussed above, $\psi_N(\lambda)$ becomes the
Baker-Akhiezer function $\psi(\lambda,t)$ in the double-scaling
limit. Then the exact, double-scaled resolvent is\foot{As is usual
when defining the continuum resolvent, one might have to impose a
cutoff on the integral at $-\Lambda$; this does not affect the
arguments below.}
\eqn\resdbl{ R(x) = \int_{-\infty}^{\infty}
{\rho_{\hbar}(\lambda)\over x-\lambda} } with
\eqn\rhodbl{
\rho_{\hbar}(\lambda) = A\hbar^2\big(\partial_\tau
\psi(\lambda,t)\partial_\lambda\psi(\lambda,t)-\psi(\lambda,t)\partial_\tau\partial_\lambda\psi(\lambda,t)\big)
} where $A$ denotes some overall numerical factor which
will be irrelevant for our purposes, and $\tau$ corresponds to the
lowest-dimension coupling as below \QPonfzzt. From the
expression for the double-scaled resolvent, it is clear that we
can think of $\rho_{\hbar}(\lambda)$ as defining the quantum
eigenvalue density.

Since the Baker-Akhiezer function is an entire function of
$\lambda$, the resolvent will be everywhere analytic, except along
the real axis, where it suffers from a discontinuity
\eqn\disc{ R(x+i\epsilon)-R(x-i\epsilon) = 2\pi i \rho_{\hbar}(x),
\qquad x\in \Bbb R } Contrast this with the classical resolvent,
which is discontinuous only along a semi-infinite cut. The
discontinuity \disc\ suggests that we define two resolvents,
$R_+(x)$ and $R_-(x)$, which are obtained by analytically
continuing $R(x)$ through the real axis from either the upper half
plane or the lower half plane, respectively. Explicitly, we define
\eqn\respm{ R_{\pm}(x) = \int_{C_{\pm}}{\rho_{\hbar}(\lambda)\over
x-\lambda} } where the contour $C_{+}$ ($C_{-}$) travels below
(above) $x$ and satisfies the same boundary conditions at infinity
as the original contour in \resdbl. Then we have \eqn\respmdblrel{
 R(x) = \cases{ R_+(x) & for $\Im x>0$\cr
                 R_-(x) & for $\Im x<0$\cr
                 }
 }
and also
\eqn\respmdblrelII{
R_+(x)-R_-(x) = 2\pi i \rho_{\hbar}(x)
}
for all $x\in \Bbb C$. Given the definition \respm, it is clear
that both $R_{\pm}(x)$ are entire functions of $x$.

Finally, let us consider the classical limit $\hbar\to 0$. In this
limit, the resolvent must reduce to the classical resolvent, which
solves the factorized loop equation and has a semi-infinite branch
cut along the real $x$ axis. Therefore, according to \respmdblrel,
$R_+(x)$ has the correct classical limit for $\Im x>0$, while
$R_-(x)$ has the correct classical limit for $\Im x<0$.
Analytically continuing the classical limits of either $R_+(x)$ or
$R_-(x)$, we find the branch cut and the second sheet of the
Riemann surface. Note that it is essential   {\it first} to take
the classical limit (i.e. drop the nonperturbative corrections),
and only then to analytically continue the resolvent. Otherwise,
we will not find the second sheet, since $R_\pm(x)$ are both
entire functions of $x$.

To summarize, we have seen that it is impossible to
define globally the quantum resolvent $R(x)$, due to the discontinuity on
the real $x$ axis. Instead, we can define through analytic
continuation {\it two} resolvents $R_{\pm}(x)$, both of which are
entire in the complex plane. So for the resolvent, just as for the
determinant, the Riemann surface disappears at $\hbar\ne 0$ and is
replaced with the complex plane. To recover the Riemann surface,
we must {\it first} take the classical limit of the resolvents,
and {\it then} analytically continue.

\subsec{More on the FZZT partition function}

We have seen above how in the double-scaling limit, all of the
observables involving the FZZT brane reduce to products,
derivatives, and integrals of a single quantity, the FZZT
partition function $\psi(x,t)$. Thus it makes sense to study this
object in more detail. Although we do not have a general formula
for $\psi(x,t)$ (see below however, where we study the example of
the Gaussian matrix model), we extracted its asymptotic behavior
at small $\hbar$ in \clasw\ using worldsheet methods. Here we
would like to rederive the WKB approximation
\eqn\heuristic{
 \psi\approx \Psi_{cl}(z,t) = (\partial_z x(z,t))^{-1/2}
 e^{\int_{x_0}^{x(z,t)} y(x,t) dx/\hbar}
 }
starting from a completely different point of view, namely the
fact that the FZZT partition function is a Baker-Akhiezer function
of the KP hierarchy. We must demonstrate that \heuristic\
satisfies \QPonfzzt\ in the $\hbar\to 0$ limit. This was first
shown in \MooreMG\ for $p=2$ (together with the interpretation in
terms of Riemann surfaces). We now give a simpler, but equally
rigorous, proof of this result.

The first step in the proof is to act on $\Psi_{cl}$ with
$d=\hbar\partial_\tau$ (at fixed $x$).  To leading order in $\hbar$, this gives
 \eqn\wkbII{ {d\Psi_{cl} \over
 \Psi_{cl}}  = \int_{x_0}^{x}\partial_\tau y\big|_x dx+\CO(\hbar)
 }
We can simplify this by writing $y=y(x(z,t),t)$ and converting the
derivative at fixed $x$ to one at fixed $z$:
\eqn\wkbIII{\eqalign{ \int_{x_0}^{x}\partial_\tau y\big|_x dx\ &=
\int_{x_0}^x \Big(\partial_\tau y\big|_z -
\partial_x y\big|_\tau \partial_\tau x\big|_z \Big)dx\cr
 &=
 \int_{z_0}^z \Big(\partial_\tau y\big|_z \partial_z x\big|_\tau -
 \partial_z y\big|_\tau \partial_\tau x\big|_z \Big)dz
}}
where we have used $\partial_z y\big|_\tau = \partial_x
y\big|_\tau \partial_z x\big|_\tau$ in the second equation. We
recognize the integrand in the second equation to be the Poisson
bracket of $x$ and $y$. Using the freedom to shift $z_0\to 0$,
together with the fact that $x$ and $y$ are given by
\eqn\QPresult{
x(z,\tau) = Q(d=z,\tau)\big|_{\hbar=0},\qquad y(z,\tau) =
P(d=z,\tau)\big|_{\hbar=0}
 }
and must therefore satisfy the genus zero string equation
\eqn\genuszerostreqn{
\{x,y\} = \partial_\tau x\partial_z y-\partial_\tau y\partial_z
x=1
}
(see appendix A for a proof of this), we conclude that
 \eqn\leadcl{
 \int_{x_0}^{x}\partial_\tau y\big|_x dx = z
 }
Using \leadcl\ in \wkbII\ we readily see that the classical
Baker-Akhiezer function $\Psi_{cl}$ is an eigenfunction of $d$
with eigenvalue $z$ in the classical $\hbar\to 0$ limit. Note
that, as mentioned in the introduction, this identity can also be
proven using worldsheet techniques for the special case of the
conformal background \SeibergNM. The advantage of the derivation
we have given here is that it is valid in every background where
the uniformizing parameter $z$ exists (i.e.\ backgrounds without
ZZ branes, in which case $\CM_{p,q}$ has genus zero).

%At the next order in $\hbar$ we need to consider several terms.
%First, using $\partial_\tau z\big|_x = {\partial_\tau
%x\over\partial_z x}$ the second term in \wkbII\ can be written as
% \eqn\parcom{\partial_\tau\left(\log \partial_z x\big|_\tau
% \right)\big|_x = {\partial_\tau\partial_z x+\partial_z^2
% x\partial_\tau z\big|_x \over \partial_z x }= {\partial_z x
% \partial_\tau\partial_z x- \partial_z^2 x\partial_\tau x \over
% (\partial_z x)^2} }
%(Except when specified explicitly, the derivatives of $x$ are
%taken when $x(z,\tau)$ is considered as a function of $z$ and
%$\tau$.)

At the next order in $\hbar$ we need to consider several terms.
The first step is to expand the operators $Q = Q_0 + \hbar Q_1 +
\cdots$, and $P = P_0 + \hbar P_1 + \cdots$,  where all
derivatives are on the right hand side. In appendix A, we show
that
\eqn\QPone{ Q_1 = {1\over2}\partial_\tau\partial_z x\big|_{z=d},\qquad
P_1={1\over2}\partial_\tau\partial_z y\big|_{z=d} } where
$x(z,\tau)$ and $y(z,\tau)$ are the classical expressions
\QPresult.
%In the expression for $Q_0$ we need to go beyond the
%classical expression \QPresult .
(Except when specified explicitly, the derivatives of $x$ are
taken when $x(z,\tau)$ is considered as a function of $z$ and
$\tau$. Thus $\p_\tau$ here is taken at fixed $z, t_{j>1}$.) When
$d$ acts on the exponent of the wavefunction \heuristic\ it gives
back $z$ as in \leadcl. In addition, we need to consider two more
terms at this order in $\hbar$. The first contribution arises from
a second derivative of the exponent. This gives a term of the form
\eqn\termextr{
 {1\over 2}  \partial_z^2 x \, \left( \partial_\tau z|_x \right)
} where the factor of ${1 \over 2} \partial_z^2 x $ comes from selecting the two
derivatives in $Q_0$ which are acting twice on the exponent of the
wavefunction, and then evaluating the rest of the derivatives
using the classical result.  The second term appears when the
derivatives of $Q$ act on the prefactor of \heuristic. This leads
to a term of the form \eqn\prefde{
 -{ 1 \over 2} \partial_z x \,
\left.  \partial_\tau \left(\log \partial_z x|_\tau
\right)\right|_x }
where again the factor $\partial_z x$ selects the derivative in $Q_0$ that
is acting on the prefactor of \heuristic .
So finally we obtain that
\eqn\Qaction{ (Q - x)
\Psi_{cl} = \hbar \left( {1\over 2}  \partial_z^2 x  \, \partial_\tau
z|_x - {1 \over 2}
\partial_z x \,
\left. \partial_\tau \left( \log \partial_z x|_\tau
\right)\right|_x + Q_1 \right) \Psi_{cl} +\CO(\hbar^2) } We now
use $\partial_\tau z\big|_x = - {\partial_\tau x\over\partial_z
x}$ to simplify these terms. In particular, we have
 \eqn\parcom{\partial_\tau\left(\log \partial_z x\big|_\tau
 \right)\big|_x = {\partial_\tau\partial_z x+\partial_z^2
 x\partial_\tau z\big|_x \over \partial_z x }= {\partial_z x
 \partial_\tau\partial_z x- \partial_z^2 x\partial_\tau x \over
 (\partial_z x)^2} }
 Then using \parcom\ and \QPone\    we find that
all terms of order $\hbar$ in \Qaction\ cancel.

%At the next order in $\hbar$  we need the $\CO(\hbar)$ correction to $Q$ and
%$P$. This was computed in appendix A, and we quote here the result
%\eqn\QPone{
%Q_1 = {1\over2}\partial_\tau\vert_x \partial_z\vert_\tau x,\qquad
%P_1={1\over2}\partial_\tau\vert_x \partial_z\vert_\tau y
%}
%Now we can compute
%
%\eqn\compbrk{
%\eqalign{
%Q \Psi_{cl} & = Q_0\biggl(z  -{1\over2}
% \hbar\ \partial_\tau \left(\log \partial_z x\big|_\tau\right)\big|_x,\tau\biggr) \Psi_{cl}
%+ \hbar Q_1 \Psi_{cl} + \CO(\hbar^2) \cr
%& = x \Psi_{cl} + \CO(\hbar^2) \cr}
%}
%

%With these facts in hand, we can finally compute the action of $Q$
%and $P$ on $\Psi_{cl}$. Using \wkbII, \QPresult, \leadcl,
%\parcom, and \QPone, we find after some cancellations
% \eqn\Qpsicl{ Q\Psi_{cl} = x\Psi_{cl}+\CO(\hbar^2)
% }
%

Computing the action of $P$ on $\Psi_{cl}$ takes a little more
work. Using again the results above, one can show that
%\eqn\Ppsicl{\eqalign{
%{(P-\hbar\partial_x)\Psi_{cl}\over\Psi_{cl}} &=
%{1\over2}\hbar\left(\partial_\tau\partial_z y+\partial_z^2 y
%\partial_\tau z\big|_x-\partial_z y \partial_\tau\log
%\partial_z x\big|_x- \partial_x \log\partial_z
%x\right)+\CO(\hbar^2)
% \cr
% &={1\over2}\hbar\left(\partial_\tau\partial_z y+\partial_z^2 y
% \partial_\tau z\big|_x-
% {\partial_z y\over\partial_z x}(\partial_\tau\partial_z
% x+\partial_z^2 x\partial_\tau z\big|_x) -
% {\partial_z^2 x\over (\partial_z x)^2} \right)+\CO(\hbar^2)\cr
% }}
%Using the partial differentiation rules we used above, we can
%simplify this to
 \eqn\PpsiclII{\eqalign{
 (P-\hbar\partial_x)\Psi_{cl} &= {1\over2}\hbar\left(
 {\partial_z \partial_\tau y\partial_z x
 -\partial_z^2 y
 \partial_\tau x- \partial_z y\partial_z \partial_\tau x\over
 \partial_z x}
 +{\partial_z^2 x(\partial_z y\partial_\tau x-1)\over(\partial_z
x)^2}
 \right)\Psi_{cl}+\CO(\hbar^2)\cr
}}
By applying the genus zero string equation \genuszerostreqn\ and
its derivative with respect to $z$, we see that the terms in
parentheses all cancel, confirming that
\eqn\PpsiclIV{
P\Psi_{cl} = \hbar\partial_x \Psi_{cl}+ \CO(\hbar^2)
}
This completes our proof that $\Psi_{cl}$ is indeed the
leading-order WKB approximation to the Baker-Akhiezer function.

Let us also offer the following non-trivial consistency check of
the semiclassical approximation \heuristic. This approximation
clearly suffers from a $p$-fold ambiguity, corresponding to which
branch of $y(x)$ and which value of $z$ in $x'(z)$ we choose. The
correct branch is chosen at large positive $x$ by demanding that
$y(x)$ be given by its physical sheet as $x\to +\infty$.
%(The
%physical sheet is characterized by $\sigma\to +\infty$ in the
%worldsheet description \xydef.)
For large $|x|$ in the first sheet we have
\eqn\yapprox{ y \approx 2^{{q\over p} -1}C x^{q\over p} } where
the real constant $C$  was determined in \KutasovFG. Its sign is
\eqn\signC{ \eta \equiv {\rm sign}(C) = - {\rm
sign}\left(\sin\left(q\pi/p\right)\right) } This means that up to
a power of $x$
 \eqn\psiapproxpq{\psi(x,t)\approx \exp\left(\eta \tilde C
 x^{p+q\over p} \right)\qquad (\Im x=0,\ \Re x\to +\infty) }
with $\tilde C$ real and positive.  We expect this semiclassical
approximation to be valid everywhere at large $|x|$, except on the
cut along the negative $x$ axis. But then the fact that
$\psi(x,t)$ is entire means that as we cross the cut, the
asymptotic behavior of $\psi(x,t)$ must change from
\eqn\psiapproxII{ \psi(x,t) \approx \exp\left(\eta \tilde C
x^{p+q\over p}\right) \qquad (\Im x>0,\ \Re x \to -\infty) } above
the negative real axis, to \eqn\psiapproxIII{ \psi(x,t)\approx
\exp\left(\eta \tilde C e^{-2 \pi i \left({p+q \over
p}\right)}x^{p+q\over p}\right) \qquad (\Im x<0,\ \Re x \to
-\infty) } below the negative real axis. Slightly above and below
the cut, both contributions \psiapproxII--\psiapproxIII\ are
present. In order for this to be consistent with the semiclassical
approximation, the first contribution must dominate above the cut,
while the second must dominate below the cut. Fortunately, this is
guaranteed by the sign of $\eta = {\rm sign}(C)$ \signC. It is
very satisfying to see how the semiclassical approximation, the
level crossing behavior, and the worldsheet calculation of $C$ all
fit together so consistently.

Finally, it is worth mentioning that this level crossing behavior
is an example of Stokes' phenomenon. We review Stokes' phenomenon
in appendix B, and we will discuss its implications for the
quantum target space in much greater detail in section 4.
\foot{Another argument that the Baker-Akhiezer function must
exhibit Stokes' phenomenon, based on the behavior of eigenvalue
distributions, was given in sec. 3.8 of \MooreCN. This argument is
related to work of F. David on nonperturbative stability \DavidSK\
and is also in accord with the discussion of nonperturbative
stability at the end of section 5 below.  }

\newsec{An Example: The Simplest Minimal String Theory and its FZZT Brane}

\subsec{FZZT correlators and the Kontsevich model}

Here we will illustrate the ideas of the previous sections using
the example of the $(p,q)=(2,1)$ model, also known as topological
gravity. Since this theory is dual to the Gaussian matrix model,
it allows us to make quite explicit some of the general formulas
derived above. Along the way we will encounter a new point of view
on the relationship between the Kontsevich matrix model and the
double-scaling limit of matrix integrals.

The $(2,1)$ model is represented in the matrix model by the
integral
 \eqn\twoone{\CZ(g)=\int dM\  e^{- {1\over g} \Tr M^2}}
with $M$ an $N\times N$ hermitian matrix. An FZZT D-brane
insertion is represented by the integral
 \eqn\dgau{\langle \det(x-M)\rangle ={1\over \CZ(g)}\int dM\ \det(x-M)
 e^{-{1\over g} \Tr M^2}}
Using \detrew, we can write this as an integral over the matrix
$M$ and $N$ fermions $\chi_i$. Then we can easily perform the
Gaussian integral over $M$ in to find the effective theory of the
fermions
 \eqn\dgaufg{\langle \det(x-M)\rangle=\int  d\chi d\chi^\dagger
 e^{-{g\over 4} ( \chi^\dagger \chi)^2 + x \chi^\dagger
 \chi}.}
Note that we started in \twoone\ with $N^2$ degrees of freedom,
the entries of $M$. After gauge fixing they are reduced to the
eigenvalues of $M$, whose number is $N$.  Now we have order $N$
fermions, but their effective theory -- which is still invariant
under $U(N)$ -- depends only on a single variable
$\chi^\dagger\chi$. To make it more explicit we replace \dgaufg\
with
 \eqn\dgaufgs{\langle \det(x-M)\rangle=\sqrt{1\over g\pi}\int
 ds d\chi d\chi^\dagger \
 e^{-{1\over g} s^2 + (i s+x)  \chi^\dagger \chi}= \sqrt{1 \over g\pi}\int  _{-\infty}^{+\infty}
 ds\ (x+is)^N e^{-{1\over g} s^2 }}
and view $s$ as an effective degree of freedom on the FZZT brane.
The final expression as an integral over $s$ is similar to the
starting point \dgau.  The matrix $M$ is replaced by a single
variable $s$ and the determinant is replaced with $(x+is)^N$.

We recognize the RHS of \dgaufgs\ as the integral representation
of the Hermite polynomials:
 \eqn\inHermite{\langle \det(x-M)\rangle=\left({g\over 4 }\right)
 ^{N\over 2} H_N\left(x \sqrt{1 \over
 g}\right)}
Since these are the orthogonal polynomials of the Gaussian matrix
model, this confirms explicitly in this example the general result
\fzztpart.

It is trivial to generalize this discussion to $n$ FZZT branes.
The partition function of $n$ FZZT branes is given by
\eqn\fzztcorrex{
\langle \det(X\otimes I_N - I_n\otimes M)\rangle = {1\over
\CZ(g)}\int dMd\chi d\chi^\dagger\ e^{-{1\over g} \Tr
M^2+\chi^{\dagger}(X\otimes I_N-I_n\otimes M)\chi}
}
with $X$ an $n\times n$ matrix and $\chi_{aj}$,
$\chi_{aj}^\dagger$ fermions transforming in the bifundamental
representation of $U(n)\times U(N)$. Integrating out $M$ and
integrating back in an $n\times n$ matrix $S$, we find (after
dropping an overall factor)
\eqn\fzztcorrexII{
\langle \det(X\otimes I_N - I_n\otimes M)\rangle = \int dS\
\det(X+iS)^N e^{-{1\over g} \Tr S^2}
}

In the large $N$ limit with $g\sim 1/N$, the eigenvalues of $M$
become localized in the interval $(- \sqrt{2},\sqrt 2)$ along the
real axis. The double-scaling limit then corresponds to zooming in
on the end of the eigenvalue distribution, while simultaneously
bringing the two saddles of \dgaufgs\ together. For example, for
$n=1$ the double-scaling limit of the FZZT partition function
\dgaufgs\ is
 \eqn\dslgauss{ x\rightarrow
 \sqrt{2}\left(1+{1\over2}\epsilon^2 \hbar^{-2/3}x\right),\quad
 N g \rightarrow 1-\epsilon^2\hbar^{-2/3}\tau, \quad s\to
 {1\over\sqrt{2}}\left(i - \epsilon s\right), \quad N\rightarrow
 \epsilon^{-3} }
with $\epsilon\to 0$. Here $\tau$ is the lowest-dimension coupling
in the continuum theory. Then \dgaufgs\ becomes (after dropping
overall numerical factors) \eqn\dslgauss{ \psi(x,t) = e^{-
x^2/2g}\langle \det(x-M)\rangle\rightarrow
\int_{-\infty}^{\infty}e^{{1\over3}i s^3+ i\hbar^{-2/3}
(x+\tau)s}\,ds } We recognize this as the Airy integral; therefore
the FZZT partition function is simply \eqn\fzztex{ \psi(x,t) =
Ai\big((x+\tau)\hbar^{-2/3}\big) } There are a few things to note
about this result.

\lfm{1.} The FZZT partition function \fzztex\ clearly satisfies
\eqn\QPex{
Q\psi = x\psi,\qquad P\psi = \hbar\partial_x \psi
}
with $Q$ and $P$ given by
\eqn\QPgauss{
Q=d^2+\tau,\qquad P=Q^{1/2}_+=d
}
(These operators obviously satisfy the string equation
$[P,Q]=\hbar$.) This confirms, in this example, that the FZZT
partition function is the Baker-Akhiezer function of the KP
hierarchy.

\lfm{2.} The Airy function \fzztex\ is an entire function in the
complex $x$ plane. On the real axis, it is oscillatory for $x\le
-\tau$ (where classically there is a cut) and decays exponentially
for $x>-\tau$. Therefore, although there appear to be two FZZT
branes with the same $x$ semiclassically (corresponding to the
different sheets of $\CM_{2,1}$), we see that the fully
nonperturbative FZZT branes depend only on $x$. The Riemann
surface disappears nonperturbatively and is replaced with only its
physical sheet.

\lfm{3.} There are, of course, two linearly independent solutions
to the equations \QPex. The other solution is the Airy function
$Bi$. We see that it does not correspond to the physical FZZT
partition function. Indeed, this solution behaves badly in the
semiclassical regime $x\to +\infty$, where it grows exponentially.
In terms of \dslgauss\ $Bi$ corresponds to another integration
contour.

\medskip

Now consider the analogous double-scaling limit for the general
FZZT correlator \fzztcorrexII. In this limit, we find
 \eqn\fzztcorrexIII{ e^{- \Tr X^2/2g}\langle \det(X\otimes I_N -
 I_n\otimes M)\rangle \to \int dS\ e^{\Tr\left(i S^3/3+
 i\hbar^{-2/3} (X+\tau)S\right)}
 }
which is, of course, the $n\times n$ Kontsevich model (for a
review of the Kontsevich model and topological gravity, see e.g.\
\DijkgraafQH). Through the use of the fermions, we have obtained a
rather direct route from the Gaussian matrix model to the
Kontsevich model. We also see quite explicitly how the matrix $S$
of the Kontsevich model is the effective degree of freedom
describing open strings stretched between $n$ FZZT branes, an
insight obtained in \GaiottoYB.

Note that we can also think of the FZZT correlator \fzztcorrexIII\
as a perturbation around the closed-string background
corresponding to $(p,q)=(2,1)$. This is a trivial statement at
finite $N$: it simply means that the insertion of determinants at
positions $x_1, \dots, x_n$ is equivalent to a certain shift in
the matrix model potential. This shift can be obtained by writing
$\det(x_i-M)$ as $e^{\Tr\log(x_i-M)}$ and expanding the logarithm
at large $x_i$.
%changing the matrix model potential from $\Tr V(M)$ to
%$\Tr V(M)+ t_k \Tr M^k $ with $t_k = { 1 \over k} \Tr ( X^{-k} )$
%[REFERENCE].
In the continuum limit, essentially the same story holds, except
that now we must expand $\log(x-M)$ in the basis of scaling
potentials $W_k(M)$ for a single cut model. For a cut between
$-\sqrt{2}$ and $\sqrt{2}$, these potentials take the form
\GrossVS
\eqn\formpot{
W'_k(M) = (2 k +1)2^{(k-1)/2} \left[
(M-\sqrt{2})^k\left(1+{2\sqrt{2}\over M-\sqrt{2}}\right)^{1/2}
\right]_+
}
where $[\ ]_+$ indicates that we expand in powers of $1/M$ and
keep only positive powers of $M$. Thus as $N\to \infty$, each
determinant insertion $\det(x-M)$ can be viewed as a
modification of the potential
\eqn\potshift{
\Tr\ V'(M)\to \Tr\ V'(M) + \sum_{k=0}^{\infty} t_{2k+1} \Tr\
W_k'(M)
}
where the couplings are given by
\eqn\tildetk{
t_{2k+1} = {2^{-(k-1)/2}\over 2k+1}{(x-\sqrt{2})^{-k-1} }\left(
1+{2\sqrt{2}\over x-\sqrt{2}}\right)^{-1/2}
}
This formula for $t_{2k+1}$ can be verified by, e.g.\ writing the
scaling potentials \formpot\ as contour integrals around infinity
and then performing the sum \potshift\ explicitly.

In the double scaling limit we zoom in on $x \sim \sqrt{2}$ as
in \dslgauss\ (we set $\hbar=1$ and $\tau=0$ for simplicity).
Summing over
$i=1,\dots, n$, we find that the $t_k$'s reduce to
\eqn\tildetkdbl{
t_{2k+1} = {1\over 2k+1}\sum_{i=1}^n x_i^{-k-1/2}
}
Therefore the $(2,1)$ model with $n$ FZZT branes can be thought of
as the closed-string background with the couplings \tildetkdbl\
turned on.

To  identify properly the precise value of the closed-string
partition function, we must also take into account the fact that
$\CZ_{\rm closed}(t) \to 1$ as $t\to 0$ (which is the same as
$x_i\to \infty$). In this limit, with $x_i \to + \infty$, the
double-scaled FZZT correlator reduces to the WKB approximation of
the matrix Airy integral \fzztcorrexIII. Thus we must divide by
this quantity to extract the closed-string partition function. In
other words, we have shown that in the double-scaling limit,
\eqn\fzztcorrexIV{ e^{- \Tr X^2/2g}\langle \det(X\otimes I_N -
 I_n\otimes M)\rangle \to C(X)\CZ_{\rm closed}(t)
 }
with
 \eqn\CX{
 C(X) = e^{-2\Tr X^{3/2}/3}\int dS\ e^{-\Tr \sqrt{X} S^2}
 }
Equating \fzztcorrexIII\ and \fzztcorrexIV, (and setting
$\hbar=1, \tau=0$) we arrive at the
relation
\eqn\kontsevichII{
\CZ_{\rm closed}(t) = {\int dS\ e^{\Tr\left(i S^3/3+
 i X S+2  X^{3/2}/3\right)}\over \int dS\ e^{-\Tr \sqrt{X} S^2}}
}
Note that by shifting the $S$ integral in the numerator, we can
rewrite this as
\eqn\kontsevichIII{
 \CZ_{\rm closed}(t) = {\int dS\ e^{\Tr\left(i S^3/3
 -Z S^2\right)}\over \int dS\ e^{-\Tr Z S^2}}
 }
with
\eqn\ZXdef{
Z = \sqrt{X}
}
Equation \kontsevichIII\ is the way the relation between the
finite $n$ Kontsevich model and topological gravity is usually
stated. Here we have rederived this fact directly from
double-scaling Gaussian matrix model.

There are a few interesting things to note in our derivation.
First, the normalizing factor $C(X)$, being the WKB approximation
to the (matrix) Airy function, is not an entire function of the
eigenvalues of $X$. For instance, it has the simple form
$(2\sqrt{\pi}x)^{-1/4} e^{-2 x^{3/2}/3}$ when $n=1$, and this
clearly has branch cuts in the complex $x$ plane.\foot{Note that
this is consistent with the asymptotic expansion at large $x$ of
our general result \clasw.} This explains why the usual relation
\kontsevichIII\ between the closed-string partition function and
the Kontsevich integral suffers from branch cuts as a function of
(the eigenvalues of) $X$. On the other hand, we see that the
combination $C(X)\CZ_{\rm closed}(t)$, being the matrix Airy
integral, is an entire function of $X$, even though the separate
factors are not.

The second point worth mentioning is the interpretation of the
quantity $Z=\sqrt{X}$ (or $Z=\sqrt{X+\tau}$ for nonzero
cosmological constant) that emerged naturally in our derivation.
This quantity also featured in the work of \GaiottoYB, where it
corresponded to the boundary cosmological constants of $n$ FZZT
branes. In order to compare with the results of \GaiottoYB, one
needs to keep in mind that in \GaiottoYB\ the Liouville coupling
constant was taken to be $b^2=1/2$, while here we are assuming
$b^2=2$. Thus, their boundary cosmological constant is equal to
our dual boundary cosmological constant $ \tilde \mu_B = \sqrt{
\mu_B + \tau} = \mu_B|_{there}$.
% $\mu_B$ is we called $x=\mu_B$ here is the dual boundary
%cosmological constant $\tilde \mu_B = \sqrt{x+\tau}$ from the
%point of view of \GaiottoYB.
%In other words, for
%us $x = \mu_B$, while for \GaiottoYB\ $ z_{\rm there} = \sqrt{ x +
%\tau} = \tilde \mu_B $. More precisely, the $(2,1)$ model
%corresponds to $b^2 =2$ in \xydef .
With our definitions, the FZZT brane labelled by $\mu_B$ is the
one corresponding to $\det(\mu_B-M)$ in the double-scaled Gaussian
matrix model. This corresponds to treating  the worldsheet
boundary interaction $ \mu_B e^{b \phi}$ as a non-normalizable
operator. From the point of view of Liouville theory with $b^2 =2$
it is more natural to consider FZZT branes as a function of
$\tilde \mu_B$, which corresponds to treating the worldsheet
boundary interaction $ \tilde \mu_B e^{ { 1 \over b} \phi}$ as a
non-normalizable operator. The expectation values of the FZZT
branes with these two choices are related, at the classical level,
by a Legendre transform. When we quantize open strings on the FZZT
brane it looks like we have a choice of which operator to fix.
These two choices amount to different quantization prescriptions
for the open strings, analogous to the different choices in other
AdS/CFT examples \KlebanovTB. It seems that the open string field
theory of \GaiottoYB\ corresponds to considering  fluctuations of
$\tilde \mu_B$. This  ends up performing a Fourier transform
between the result for fixed $\tilde \mu_B$, which is a simple
exponential, and the result at fixed $\mu_B$, which is given by
the Airy integral.

\subsec{The quantum resolvent in the Gaussian matrix model}

The Gaussian matrix model also provides a good setting for the
discussion of the quantum resolvent in section 2.4. Substituting
\fzztex\ into \rhodbl, we find the quantum eigenvalue density
\eqn\rhogauss{
\rho_{\hbar}(\lambda) =
\hbar^{1/3}Ai'((\lambda+\tau)\hbar^{-2/3})^2-\hbar^{-1/3}(\lambda+\tau)Ai((\lambda+\tau)\hbar^{-2/3})^2
}
One can check that this is everywhere positive on the real axis.
Using the asymptotic expansions
\eqn\BAasymptotic{\eqalign{
 &Ai(x)\sim \cases{{1\over 2\sqrt{\pi}x^{1/4}}e^{-2/3x^{3/2}}&\quad
 $|\arg(x)| < {\pi}$\cr
 {1\over\sqrt{\pi}(-x)^{1/4}}\sin\left({\pi\over4}+{2\over3}(-x)^{3/2}\right)
 &\quad $\arg(x)= \pi $}
 }}
we see that the classical limit of the eigenvalue density on the
real axis is as expected:
\eqn\rhogausscl{
 \lim_{\hbar\to0}\rho_{\hbar}(\lambda) = \cases{{\hbar\over 8\pi
 (\lambda+\tau)}e^{-{4\over3\hbar}(\lambda+\tau)^{3/2}} &\quad $\lambda>-\tau$\cr
 {\sqrt{-(\lambda+\tau)}\over\pi}&\quad $\lambda<-\tau$\cr}
 }

Now consider the quantum resolvent. Since $\rho_{\hbar}$ is
positive on the real axis, the quantum resolvent \resdbl\ is
indeed discontinuous across the entire real axis. Combining
\respmdblrelII\ with \BAasymptotic, we see that the resolvents
$R_+$ and $R_-$ only differ by a small, nonperturbative amount in
a wedge around the positive real axis (we now set $\tau=0$ for
simplicity).
\eqn\resgauss{
R_+(x) - R_-(x) \sim {i\hbar\over 4
 x}e^{-{4\over3\hbar}x^{3/2}},\qquad |\arg(x)| \le {\pi\over 3}
 }
Let us call this wedge region I. By the same argument, the
resolvents differ by a {\it large} amount in the wedge $
{\pi\over3}< |\arg(x)| \le \pi$, which we will call region II. In
other words, the small nonperturbative quantity \resgauss\ in
region I becomes a large nonperturbative quantity in region II. In
region II, the resolvent $R_+$ has the correct classical limit in
the upper half plane, while the resolvent $R_-$ has the correct
classical limit in the lower half plane.

To find the second sheet of the Riemann surface, we must first
take the classical limit of $R_+$ ($R_-$) in the union of region I
and the upper (lower) half plane. Only by dropping the
nonperturbative corrections does the branch cut at $x<0$ appear.
Then we can analytically continue through this cut to find the
second sheet.

\newsec{The Effective Theory on the Brane}

We have seen in the previous two sections how the Riemann surface
disappears nonperturbatively, with the FZZT partition function
being an entire function of the complex $x$ plane. Here we would
like to understand this nonperturbative modification in more
detail, from the point of view of the effective theory on the FZZT
brane. We will limit ourselves to the simplest case of $(2,1)$, in
which case the effective theory \fzztcorrexIII\ on $n$ FZZT branes
is the $n\times n$ Kontsevich model \GaiottoYB. For simplicity, we
will consider only the case $n=1$.

The semiclassical approximation of $\hbar\to 0$ corresponds to the
saddle-point approximation. For the FZZT partition function
\dslgauss, there are two saddle points in the $s$ integral,
located at
\eqn\saddles{
\langle s\rangle  = \pm \hbar^{-1/3}\sqrt{-x}
}
(For simplicity we set $\tau=0$.) Therefore, there are two
distinct branes for each $x$, semiclassically. The moduli space of
branes becomes a double cover of the $x$ plane, as we saw in the
introduction.

Contrast this now with the quantum theory. Here we must integrate
over $s$; i.e.\ we must study the quantum dynamics of the theory
on the brane. The subleading saddles in the integral over $s$ are
{\it instantons} in the theory on the brane. As is always the case
with instantons, one must sum over them in some prescribed
fashion. The result of this process is that the exact,
nonperturbative FZZT partition function becomes an entire function
of $x$. For $(2,1)$ it is the Airy function \fzztex.

The crucial point is that instead of exhibiting monodromy around
$x=\infty$, the FZZT partition function now exhibits what is known
as Stokes' phenomenon. As we discussed in the introduction, this
is the phenomenon whereby the analytic continuation of the
asymptotics of a function in one region does not correctly
reproduce the asymptotics of the function in another region. (See
also appendix B for a brief review of Stokes' phenomenon,
summarizing \Berry.)

The Airy function is actually a paradigmatic example of Stokes'
phenomenon. In the region $x\to +\infty$, the Airy function is
given approximately by
 \eqn\airyasymp{ Ai(x) \sim {1\over 2\sqrt{\pi}x^{1/4}}
 e^{-{2\over3}x^{3/2}} }
Attempting to analytically continue the asymptotics
counterclockwise around large $x$ to $x\to -\infty$, one would
find $Ai(x)\sim e^{+{2\over3}i(-x)^{3/2}}$ there. However, the
correct asymptotics of the Airy function on the negative real axis
is actually
 \eqn\airyasympII{ Ai(x) \sim {1\over \sqrt{\pi}(-x)^{1/4}}
 \sin\left({\pi\over 4} + {2\over3}(-x)^{3/2}\right) }
i.e.\ it is a linear combination of the two saddles \saddles.

The reason this happened is that as we varied $x$ from
$x=+\infty$ to $x=-\infty$, we crossed a Stokes' line at
 \eqn\stokeslines{ \arg(x)=\pm {2\pi \over 3} }
Recall from the introduction that Stokes' lines are the places
where various saddle-point contributions to an integral appear and
disappear. In our example, one can see this by starting from the
negative real axis, where according to \airyasympII, both saddles
contribute to the Airy integral. Upon crossing the Stokes' lines
\stokeslines, however, the subdominant saddle disappears entirely
from the asymptotic expansion, until one reaches the positive real
axis, where the function is given by \airyasymp. The Stokes' lines
occur at precisely the points where the disappearing saddle is
most subdominant.

We can also describe the effect of Stokes' lines in a slightly
different way: the presence of Stokes' lines implies that in some
regions, certain saddle points might not contribute at all to the
integral. Therefore, the naive procedure of just summing over all
the saddle points is not valid here. A case in point is again the
$x\to +\infty$ asymptotics of the Airy function \airyasymp. There
we see that the function is dominated by just one saddle. The
other saddle clearly does not contribute; if it did, it would
contribute an exponentially {\it increasing} contribution to the
Airy function at $x\to +\infty$.

Let us conclude this section by briefly summarizing two general
lessons we can learn from this example.

\lfm{1.} The classical saddle-point approximation is certainly
valid, but in the exact quantum answer, we might need to sum over
saddles. Because of Stokes' phenomenon, not all the saddles
necessarily contribute in various asymptotic regions. Even the
dominant saddle sometimes does not contribute.\foot{ The relation
between the various sheets of the Riemann surface and the exact
answer was explored also in \DijkgraafVP  , where the proposal
seems to be to sum over all saddles.}

\lfm{2.} The quantum target space (the complex $x$ plane) differs
significantly from the classical target space (the Riemann
surface). The various unphysical sheets of the Riemann surface
disappear, also because of Stokes' phenomenon. In a wedge around
the erstwhile branch cut, what were classically interpreted as the
unphysical sheets lead to exponentially small corrections to the
exact, quantum answer.

\newsec{Comments on Other Backgrounds}

In section 2, we extracted the asymptotics of the FZZT partition
function $\psi(x,t)$ at large positive $x$ and small $\hbar$ for
general $(p,q)$. Combining this with the fact that $\psi(x,t)$ is
an entire function of $x$, we argued that $\psi(x,t)$ exhibited
Stokes' phenomenon. Thus, we expect that our conclusion above
about the disappearance of the Riemann surface is generally true
and that this phenomenon comes about in a way similar to what we
saw in the $(p=2,q=1)$ model. Certain regions in the unphysical
sheets lead to small nonperturbative corrections to the
semiclassical answer in the physical sheet. Meanwhile, other
regions in the unphysical sheets do not contribute such effects.

%%%
However, a true generalization of our analysis from $(2,1)$ to other
values of $(p,q)$ requires clarification of two issues: first,  the role of the ZZ
branes, which exist for higher $(p,q)$ but not for $(2,1)$,  and second,
the overall nonperturbative consistency of models with higher
$(p,q)$.

First, let us discuss the ZZ branes. Recall that at the classical
level, the number of background ZZ branes is measured by the
$A$-periods of the one-form $y dx$ on the Riemann surface
$\CM_{p,q}$ \SeibergNM:
\eqn\ZZNi{
\oint_{A_i} ydx=N_i \hbar
}
where $N_i$ is the number of ZZ branes of type $i$. In the
simplest backgrounds without ZZ branes the $A$-periods vanish and
the surface degenerates to a genus zero surface. The moduli of
$\CM_{p,q}$ fall into two classes \refs{\SeibergNM,\KutasovFG}.
Moduli which preserve the $A$-periods correspond to closed string
backgrounds. Moduli which change the values of these periods arise
only when ZZ branes are added.

It is clear that this picture must be modified in the exact
quantum theory.  Even without a nonperturbative definition of the
theory it is clear that the periods $ \oint_{a_i} ydx=N_i \hbar $
are quantized when $\hbar \not=0$ and therefore they cannot change
in a continuous fashion by varying moduli. But in order to
understand the exact nonperturbative theory, we need a definition
of the theory which goes beyond the worldsheet expansion.  We take
the double scaled matrix model to be this definition.

In the matrix model, the ZZ branes represent eigenvalue
instantons, corresponding to sub-leading saddles of the matrix
integral where some of the eigenvalues are away from the cut.
Thus, unlike the classical theory, which is characterized by fixed
values of the integers $N_i$, in the exact theory we must sum over
the $N_i$. This sum over ZZ branes is automatically incorporated
in the exact, nonperturbative matrix integral. In the end, the
exact answer is characterized only by the closed string
backgrounds.

The second issue concerns the nonperturbative existence of the
$(p,q)$ models. It is well known that certain values of $(p,q)$
(e.g.\ $(p,q)=(2,3)$, corresponding to pure gravity) do not exist
nonperturbatively \refs{\BrezinVD\DavidGE\DavidSK-\DouglasXV,
\MooreCN,\MooreMG}.  This happens when the double-scaled matrix
model potential is not bounded from below.  We can study this
problem in the continuum using the FZZT partition function and its
relation to the effective potential
\refs{\KlebanovWG,\SeibergNM,\KutasovFG}
 \eqn\psiveff{\psi(x,t) \approx e^{-V_{eff}(x)/2\hbar}}
For large $|x|$ away from the negative real axis we have from
\psiapproxpq
  \eqn\psiapproxpqa{ {1\over2\hbar}V_{eff}(x) \approx -\eta \tilde C
  x^{q+p\over p}}
Therefore, the effective potential is bounded from below only when
\KutasovFG
\eqn\bounded{
\eta = - {\rm sign}(\sin(q\pi/p))<0
}
For example, in the $(p=2,q=2l-1)$ models \bounded\ is satisfied
only for $l$ odd, and it is never true in the unitary models with
$(p,q=p+1)$.

Notice that for the nonperturbatively consistent models,
$\psi(x,t)$ vanishes as $x \to +\infty$. This reflects the fact
that the eigenvalues are not likely to be found there. This also
specifies boundary conditions for the differential equations
\QPonfzzt\ satisfied by $\psi$, and these boundary conditions are
sufficient to  determine uniquely $\psi(x,t)$. Conversely, in the
nonperturbatively inconsistent models the semiclassical value of
$\psi(x,t)$ diverges as $x \to +\infty$. Thus there is a
nonperturbative ambiguity in the definition of $\psi(x,t)$
corresponding to the freedom to add small exponential corrections
to the dominant contribution \psiveff. So we see how the
nonperturbative problems of these models, which are associated
with the unboundedness of the potential, manifest themselves here
in the ambiguity of defining the FZZT partition function
$\psi(x,t)$.

\newsec{Summary and Discussion}

We have explored the relation between the semiclassical geometry
seen by the FZZT branes in minimal string theory and the exact
results as computed by the matrix model. We have seen that the
various sheets of the Riemann surface correspond to different
saddle points of the effective theory on the FZZT brane. For some
ranges of the value of the boundary cosmological constant all the
saddles contribute to the answer, while for some other ranges only
a subset of all saddles contribute. The precise matrix model
definition of the theory tells us which saddles contribute and
which do not contribute. So the Riemann surface, which plays a
crucial role in the perturbative analysis of the model, suffers
drastic modifications when we consider the full nonperturbative
aspects of the theory.

We have also given a quick derivation of the relation between the
Kontsevich matrix model and the ordinary double scaled matrix
model, clarifying the relation between the various open string
descriptions of the theory. The Kontsevich model arises as the
effective theory on the FZZT branes \GaiottoYB\ after integrating
out all the open strings corresponding to the ZZ branes. The
degrees of freedom of the Konsevich model are, roughly speaking,
``mesons" made out of the fermionic strings stretched between FZZT
and ZZ branes \KutasovFG.

We have discussed in detail the simplest $(2,1)$ model, but we
also argued on general grounds that the disappearance of the
Riemann surface due to Stokes' phenomenon is a feature of all of
the $(p,q)$ models.

Our paper was partially motivated by the discussion of physics
behind the horizon in \FidkowskiNF  . We have a somewhat similar
problem. The Riemann surface is analogous to spacetime and the
second sheet is analogous to the region behind the horizon. The
FZZT brane observables could be viewed as probes of the spacetime
geometry. Note that the parameter $x=\mu_B$ is controlled in the
boundary region ``outside" the horizon. By analytically continuing
semiclassical answers we get to explore the second sheet of the
Riemann surface, much in the same way that the region behind the
horizon is explored in \FidkowskiNF  . The black hole singularity,
where quantities diverge, is somewhat analogous to the $x \to
\infty$ region of the second sheet, where again the expectation
values of the analytically continued FZZT branes diverge. In the
exact answer, however, this saddle point ceases to contribute
before its value becomes very large. In both cases, the
holographic theory tells us that there are no divergences in this
region. We expect that further analysis of these simple exactly
solvable examples might yield interesting general lessons for how
to think about quantum gravity in higher dimensions.

Another motivation for our work was the resemblance, at least at
the perturbative level, between the minimal string theories and
the topological string (see, e.g.\ the discussion of the $(p,1)$
models in \AganagicQJ). Nonperturbatively, however, the connection
is less clear. While we lack a generally accepted nonperturbative
definition for the topological string, minimal string theories
have an exact, nonperturbative formulation in terms of the dual
matrix model. Given the similarities between the two theories, it
is natural to suppose that some of the lessons from our work might
be relevant to the topological string. Let us just briefly mention
a few.

First, we have seen how the semiclassical target space (the
Riemann surface) is drastically modified by nonperturbative
effects. In the topological string, the Riemann surface is
%%%
intimately connected with the target space. The Riemann surface
is the surface $H(x,y)=0$ in ${\Bbb C}^2$. The   Calabi-Yau
is given by the equation
  $uv + H(x,y)=0$ in ${\Bbb C}^4$. This Calabi-Yau is   a
 ${\Bbb C}$-fibration over the complex
$(x,y)$ plane, and the discriminant locus of the fibration is
the Riemann surface $H(x,y)=0$.  Our results raise the question of whether in  a
proper nonperturbative definition of the topological string the
Calabi-Yau might  also be drastically modified by quantum effects.

Another striking feature of our analysis is the role of Stokes'
phenomenon. Semiclassical target space is viewed as a saddle-point
approximation to some effective theory on the brane probe.
Nonperturbatively, we must sum over different saddle-points in a
prescribed fashion. As we saw with the Airy function, the result
was that some portions of the semiclassical target space
contributed to physical observables, while others did not. It even
happened sometimes that the dominant saddle-point did not
contribute. It will be interesting to see if these phenomena play
a role in the nonperturbative topological string.

A third possible application to the topological string is the role
of the ZZ branes. These correspond to eigenvalue instantons in the
matrix model. The classical vacuum of the matrix model corresponds
to placing all of the eigenvalues into the dominant minimum of the
matrix model potential. However, in the exact answer we must sum
over all vacua (i.e.\ sum over all instantons), obtained by
filling the other critical points of the potential with any number
of eigenvalues. In the continuum limit, this means that we must
integrate over a subset of the moduli of the Riemann surface
describing the normalizable deformations due to ZZ branes. Note
that the closed-string couplings are non-normalizable
deformations, and hence we do not integrate over them. This
suggests that in the nonperturbative topological string, one
should also integrate over some of the moduli of the Calabi-Yau
(the normalizable modes) but not others.

In the context of the topological string, it has been suggested
that the Riemann surface is covered by patches and the D-branes in
different patches are related by (generalized) Fourier transform
\refs{\AganagicQJ,\DijkgraafVP}.  Comparison with the two matrix
model suggests, as mentioned in footnote 3, that the theory has
two distinct branes $\det(x-M)$ and $\det(y-\tilde M)$.  As in
\KazakovDU, these are natural in different patches on $\CM_{p,q}$
consisting of the first sheet of the $x$-plane and the first sheet
of the $y$-plane (note that these two patches do not generally
cover the whole surface).  As in \refs{\SeibergNM,\KazakovDU}, in
the classical theory these branes are related by Legendre
transform, with the boundary cosmological constant $x$ and its
dual $y$ satisfying the defining equation of $\CM_{p,q}$.  It is
likely that nonperturbatively they are related by a Fourier
transform. This can be interpreted as a relation between different
branes rather than as a relation between different patches of the
surface.

%%%
%%%
Finally, let us mention a more mathematical potential application
of our work. We have seen that the asymptotics of the
Baker-Akhiezer function \heuristic\ shows very clearly the
emergence of the classical Riemann surface through the one-form
$ydx$. It is an old idea \MooreMG\ that the full Baker-Akhiezer
function should be used to define a ``quantum Riemann surface,''
associated with the string equations $[P,Q]=\hbar$ in a way
analogous to the association of a Riemann surface to the
stationary KdV equations, in which case $[P,Q]=0$. A closely
related point is the relation of the matrix model partition
function and KdV flows to the infinite Grassmannian.  In
particular, in the free fermion interpretation of the infinite
Grassmannian one needs to introduce an operator which does not
create monodromy, (such as twistfields in conformal field theory)
but rather Stokes multipliers. Such operators, called ``star
operators'' in \refs{\MooreCN,\MooreMG},  are not at all
well-understood. It was suggested in \refs{\MooreCN,\MooreMG}\
that the point in the Grassmannian created by star operators
should define a ``theory of free fermions on a quantum Riemann
surface.'' A similar suggestion has recently been made in
\AganagicQJ. Perhaps it is a good time to revisit these issues.
%%%

\vskip 0.8cm

\noindent {\bf Acknowledgments:}

We would like to thank S.~Shenker for useful discussions. GM would
like to thank the Aspen Center for Physics for hospitality during
the completion of this paper. The research of JM and NS is
supported in part by DOE grant DE-FG02-90ER40542. The research of
GM is supported in part by DOE grant DE-FG02-96ER40949. The
research of DS is supported in part by an NSF Graduate Research
Fellowship and by NSF grant PHY-0243680. Any opinions, findings,
and conclusions or recommendations expressed in this material are
those of the author(s) and do not necessarily reflect the views of
the National Science Foundation.

%\vskip 2cm

\appendix{A}{Geometric Interpretation of the Lax Formalism}

In this appendix, we will study the Lax operators $Q$ and $P$ in
the semiclassical $\hbar\to 0$ limit. Much of this section will
consist of collecting and streamlining many facts that are
scattered throughout the literature. However, in the process of
organizing this material, several new insights will emerge.

Our main goal is to provide a geometric interpretation for $P$ and
$Q$ in terms of the Riemann surface $\CM_{p,q}$ of minimal string
theory. How this geometric interpretation is modified at $\hbar\ne
0$ is an important question. In \MooreMG, it was proposed that
%%%
by generalizing the Burchnall-Chaundy-Krichever theory of KdV flow,
phrased in terms of framings of line bundles, to
framings of a flat holomorphic vector bundle over the space of
$x,t_j$, one could define a notion of a ``quantum Riemann surface.''
It would be nice to understand the relation of this
proposal to the geometrical interpretation given below.

\subsec{A brief review of the Lax formalism}

First, let us take a moment to  recall briefly the definition of
the Lax operators $Q$ and $P$ of minimal string theory. (For a
more thorough review, see e.g.\ \DiFrancescoNW.) These operators
are a convenient way to package neatly the data (physical
correlation functions) of minimal string theory. They are
differential operators, of degree $p$ and $q$ respectively, in
\eqn\ddef{
d=\hbar \partial_{\tau}
}
where $\tau=t_1$ is the coupling to the lowest-dimension operator.
Explicitly, we have
\eqn\QPdef{\eqalign{
 Q&\propto d^{p}+{1\over2}\sum_{j=2}^{p}\{u_{p-j}(t),d^{p-j}\}\cr
 P&\propto d^{q}+{1\over2}\sum_{j=2}^{q}\{v_{q-j}(t),d^{q-j}\}\cr
 }}
where the coefficients $u_{p-j}(t)$ and $v_{q-j}(t)$ represent
various two-point functions of physical closed-string operators.
They depend on the closed-string couplings $t=(t_1, t_2, \dots)$.
For instance, \eqn\freeenergy{ u_{p-2}(t) \propto
\partial_\tau^2\log \CZ } corresponds to the ``specific heat" of the
string theory.

To   solve minimal (closed) string theory, we simply need
to solve for the dependence of $Q$ and $P$ on the closed-string
couplings $t=(t_1,t_2,\dots)$. This is done by requiring that $Q$
and $P$ satisfy the string equation
\eqn\stringeqn{
[P,Q]=\hbar
}
along with the KdV flows
\eqn\KdV{
\hbar{\partial Q\over \partial t_j} = [Q,Q^{j/p}_+],\qquad
\hbar{\partial P\over \partial t_j} = [P,Q^{j/p}_+]
}
The compatibility of the latter with the former implies that $P$
is given in terms of $Q$ by
\eqn\Psol{
P = \sum_{k\ge 1\atop k\ne 0\,\mod\,p}(1+k/p)\ t_{k+p}\ Q^{k/p}_+
}
Substituting this back into \stringeqn\ then gives a set of
coupled differential equations for the coefficient functions of
$Q$. These equations can be solved order by order in $\hbar$,
resulting in a perturbative expansion for $P$ and $Q$
 \eqn\PQexpand{\eqalign{
 Q &= Q_0(d,t)+\hbar\, Q_1(d,t)+\hbar^2\, Q_2(d,t)+\dots \cr
 P &=
 P_0(d,t)+\hbar\, P_1(d,t)+\hbar^2\, P_2(d,t)+\dots
 }}
where by convention the operators on the RHS of \PQexpand\ are
ordered such that the $d$'s are all on the right.

\subsec{The Lax operators in the semiclassical limit}

Now let us take $\hbar\to 0$ to obtain a much simpler set of
equations for the Lax operators. In this limit, the string
equation \stringeqn\ becomes
\eqn\streqnclass{
{\partial P_0\over \partial d}{\partial Q_0\over \partial
\tau}-{\partial P_0\over \partial \tau}{\partial Q_0\over \partial
d}=1
}
i.e.\ the commutator is replaced with a Poisson bracket. To see
this, note that every time $d=\hbar\partial_\tau$ acts on
something to its right, it contributes a factor of $\hbar$.
Therefore the leading order contribution to the commutator is the
Poisson bracket \streqnclass.

The solution to this equation is well-known (see e.g.\ section 4.5
of \DiFrancescoNW). It is simply
\eqn\classsol{\eqalign{
 P_0(d;t) = y(x;t)\quad \hbox{with $x=Q_0(d;t)$}
 \cr
 }}
where $y(x;t)$ is the singular part of the large $N$ matrix model
resolvent in the closed-string background labelled by $t$. Since
$(x,y)$ lie on the Riemann surface $\CM_{p,q}$, \classsol\ implies
that at $\hbar=0$,
%%%***
the simultaneous eigenvalues of $Q_0$ and $P_0$ also lie on the
same Riemann surface, i.e. \eqn\QPMpq{ (Q_0,P_0)\in \CM_{p,q} }
The fact that they can be written in the form \QPdef\ as
polynomials in $d$ implies that the eigenvalue of $d$ is the
uniformizing parameter for $\CM_{p,q}$. Thus we can write
\classsol\ as follows:
\eqn\classsolrew{ Q_0=x(z=d;t),\qquad P_0=y(z=d;t) } Thus we have
reduced the algebraic-differential problem of solving the genus
zero string equation to the geometric problem of finding the
uniformizing parameter of $\CM_{p,q}$. This problem has been
solved in various special cases. For instance, in \SeibergNM, it
was found that
\eqn\foundthat{ x(z) = T_p(z),\qquad y(z)=T_q(z) }
in the conformal background. (To keep the equations simple in this
appendix, we will rescale $y$ so as to remove the coefficient $C$.
This will have no effect on arguments below.)

 Although it is in general a nontrivial exercise to extract
from the string equation the higher order $\hbar$ corrections to
the Lax operators, it is actually easy to obtain the first order
$\hbar$ corrections $Q_1$ and $P_1$. This is because the
coefficient functions of $Q$ and $P$, being closed-string
observables, have an expansion in $\hbar^2$ (the closed-string
coupling), not $\hbar$ (the open string coupling). (Note that this
statement is only true for the particular ordering prescription we
used in defining the Lax operators \QPdef.) Thus $Q_1$ and $P_1$
arise only from the non-commutation of $d$ and the coefficient
functions. This gives
 \eqn\QPone{
 Q_1 = {1\over2}\partial_\tau\partial_z x(z,\tau)\big|_{z=d},\qquad
 P_1={1\over2}\partial_\tau\partial_z y(z,\tau)\big|_{z=d}
 }
Here we have used \classsolrew, and, as noted above, $Q_1$ and
$P_1$ are defined with the $d$'s all on the right.

Finally, we should note that the discussion of $P$ and $Q$ in this
appendix is limited to the classical backgrounds without ZZ
branes, where the surface $\CM_{p,q}$ has genus zero and a number
of pinched cycles. It will be interesting to see how to generalize
this discussion to backgrounds with ZZ branes present. Then the
pinched cycles of $\CM_{p,q}$ are opened up and the surface no
longer has genus zero. In such backgrounds, $z$ is no longer a
good uniformizing parameter, and our interpretation of $P$ and $Q$
will have to be modified accordingly.

\bigskip\noindent{\it KdV flow and deformations of $\CM_{p,q}$}

Having shown that the simultaneous eigenvalues of
$Q$ and $P$ (we will drop the subscript 0 from
this point onwards) are nothing but the coordinates $(x,y)$ of
$\CM_{p,q}$, we can now provide a geometric interpretation of the
KdV flow equations \KdV. The KdV flows tell us how to deform $Q$
and $P$ from a closed-string background $t$ to a nearby background
$t+\delta t$. This gives rise to a deformation of $\CM_{p,q}$.
Therefore, on general grounds, the genus-zero KdV flows must be
equivalent to the singularity-preserving deformations of
$\CM_{p,q}$ discussed in \SeibergNM.

We can check our claim explicitly in the conformal background.
After a lengthy calculation, whose details we will skip, one
derives the following deformations of $P$ and $Q$ from the KdV
flow equations \KdV:
\eqn\KdVfinal{
{\partial Q\over \partial\tau_{r,s}} ={1\over q}
U_{p-1}(d)\left[{T_{ps}(d)U_{qr-1}(d)\over
U_{p-1}(d)U_{q-1}(d)}\right]_-
}
and
\eqn\KdVfinalII{
{\partial P\over \partial\tau_{r,s}} ={1\over p}
U_{q-1}(d)\left[{T_{qr}(d)U_{ps-1}(d)\over
U_{p-1}(d)U_{q-1}(d)}\right]_- -{1\over p}
U_{q-1}(d)\left[{U_{qr-ps-1}(d)\over
U_{q-1}(d)U_{p-1}(d)}\right]_+
}
in the conformal background, up to an overall normalization
factor. Here $\tau_{r,s}$ is the coupling associated to the
continuum operator $\CV_{r,s}$; it is related to the matrix model
couplings $t_j$ by a linear transformation. (The change of basis
between matrix model and continuum couplings is discussed in
\MooreIR.) It is important that both \KdVfinal\ and \KdVfinalII\
are polynomials in $d$; this is required by the definition of $Q$
and $P$. Note also that the degree of the deformation to $Q$ is
always less than $p$, but there is no restriction on the degree of
the deformation to $P$.

Since the curve for $\CM_{p,q}$ in the conformal background is
\eqn\curve{
F(Q,P) = T_q(Q)-T_p(P)= 0
}
the deformation to the curve due to \KdVfinal--\KdVfinalII\ is
\eqn\curvedef{\eqalign{
    {\partial F\over \partial \tau_{r,s}} &=
 U_{q-1}(Q)U_{p-1}(d)\left[{T_{ps}(d)U_{qr-1}(d)\over
 U_{p-1}(d)U_{q-1}(d)}\right]_- -
 U_{p-1}(P)U_{q-1}(d)\left[{T_{qr}(d)U_{ps-1}(d)\over
 U_{p-1}(d)U_{q-1}(d)}\right]_-\cr
& \qquad\qquad +U_{p-1}(P)
 U_{q-1}(d)\left[{U_{qr-ps-1}(d)\over
 U_{q-1}(d)U_{p-1}(d)}\right]_+\cr
 &=
 U_{pq-1}(d)\left({T_{ps}(d)U_{qr-1}(d)-T_{qr}(d)U_{ps-1}(d)\over
 U_{p-1}(d)U_{q-1}(d)}\right)\cr
}}
In the second line, we have substituted \foundthat\ for $Q$ and
$P$ and we have used the identity
$U_{m-1}(T_n(z))=U_{mn-1}(z)/U_{n-1}(z)$. Further use of this
identity leads to
\eqn\curvedefII{
  {\partial F\over \partial \tau_{r,s}} = U_{q-1}(Q)T_{s}(Q)U_{r-1}(P)-U_{p-1}(P)T_{r}(P)U_{s-1}(Q)
}
which agrees exactly with the singularity-preserving deformations
of $\CM_{p,q}$ found in \SeibergNM. This confirms very explicitly
the equivalence between the KdV flows and the deformations of
$\CM_{p,q}$.

We should mention that for $p=2$, the equivalence of the KdV flows
and the singularity-preserving deformations of $\CM_{p,q}$ can be
seen more directly using the formulas in \refs{\MooreCN,\MooreMG}.
There it is shown, using the representation of the KdV equations
as first-order matrix equations, that one can define an
``$\hbar$-deformed" Riemann surface $y^2=F(x;t,\hbar)$ which
reduces as $\hbar\to0$ to the classical Riemann surface (what we
call $\CM_{p,q}$) of the matrix model. Here $F(x;t,\hbar)$ is a
polynomial in $x$, which depends in a complicated way on the
closed string couplings $t$. Although we will not discuss the
details here, one can show that at $\hbar=0$, the Riemann surface
reduces to
\eqn\Mpqgreg{
y^2=(x+u(t))\bigl(B(x;t)\bigr)^2
}
%%%
where $B(x,t)$ is a polynomial in $x$ as well as in the Gelfand-Dickii
potentials $R_j[u]$. (See eq. $(2.35)$ of \MooreCN.)
 The form \Mpqgreg\ shows immediately that the KdV flows are singularity-preserving
deformations of $\CM_{p,q}$, since as we change the couplings $t$,
the RHS of \Mpqgreg\ always has only one branch point at $x=-u(t)$
and singularities at the roots of $B(x;t)$.

It is interesting to contrast this with the Burchnall-Chaundy-Krichever theory of
stationary KdV flows. There the Riemann surface is obtained from simultaneous
eigenvalues of the differential operators $[P,Q]=0$. The KdV flow preserves
the Riemann surface moduli and instead is straight-line flow along the
Jacobian of the Riemann surface \segalwilson.

\bigskip\noindent{\it Instantons and the singularities of $\CM_{p,q}$}

Finally, we will discuss the connection between instantons and the
singularities of $\CM_{p,q}$. Instantons were studied using the
classical limit of the Lax formalism by Eynard and Zinn-Justin in
\EynardSG. Let us briefly review the logic of their analysis. To
leading order, an instanton corresponds to an
exponentially-suppressed perturbation $\epsilon(t)$ of the
specific heat $u(t)$ and all other physical correlation functions.
Thus in the $\hbar\to 0$ limit, we can write
\eqn\instanton{
\epsilon'/\epsilon = r\sqrt{u(t)}
}
for some constant $r$ which measures the strength of the
instanton. (The derivative in \instanton\ is with respect to the
lowest-dimension coupling $\tau$.) Since as $\hbar\to 0$ we can
ignore the $t$ dependence of $u(t)$, we might as well set
$u(t)=1$. Then \instanton\ can be written as $d\epsilon =
\epsilon(d+r)$, which implies that
\eqn\instantonrew{
f(d)\epsilon=\epsilon f(d+r)
}
for any function $f(d)$.

The next step in the analysis of \EynardSG\ is the observation
that the instanton deforms the Lax operators by
\eqn\instdef{
\delta Q = \epsilon S(d),\qquad \delta P = \epsilon R(d)
}
where $S(d)$ and $R(d)$ are polynomials in $d$ of degree $p-2$ and
$q-2$ respectively. Since this deformation must preserve the
string equation $[P,Q]=\hbar$, this leads to the following
constraint at linear order in $\epsilon$:
\eqn\instconstraint{
[P,\delta Q]+[\delta P,Q]=0
}
Substituting \instdef\ and using \instantonrew, we find
\eqn\instconstraintII{
(P(d+r)-P(d))S(d)=(Q(d+r)-Q(d))R(d)
}
This constraint must be satisfied for every $d$ and for some
constant $r$. Since $Q(d+r)-Q(d)$ and $P(d+r)-P(d)$ are degree
$q-1$ and $p-1$ respectively, but $S(d)$ and $R(d)$ are only
degree $p-2$ and $q-2$ respectively, \instconstraintII\ implies
that $Q(d+r)-Q(d)$ and $P(d+r)-P(d)$ must share a common root.
Thus there exists some $d=d_0$ where
\eqn\instimplies{
\big(Q(d_0+r),P(d_0+r)\big)=\big(Q(d_0),P(d_0)\big)
}
The authors of \EynardSG\ use \instimplies\ to solve for $r$, and
then use \instconstraintII\ to solve for $S(d)$ and $R(d)$.

With the geometric interpretation of the previous sections in
hand, we can offer some new insights into the analysis of the
instantons. The condition \instimplies\ is equivalent to the
condition that $\CM_{p,q}$ have a singularity (pinched cycle) at
the point $(x,y) = (Q(d_0),P(d_0))$. This shows that the
instantons are in one-to-one correspondence with the singularities
of $\CM_{p,q}$. It confirms in a direct way the analysis of
\SeibergNM\ and the interpretation of the ZZ branes as instantons.

In \SeibergNM, it was also argued that the period of $y\,dx$
around the $B$-cycle passing through the $(m,n)$ singularity
computes the $(m,n)$ instanton (ZZ brane) action, i.e.
\eqn\instZ{
Z_{m,n} \propto \oint_{B_{m,n}}y\, dx =
\int_{z_{m,n}}^{z_{m,n}+r_{m,n}}y(z)x'(z)dz
}
with the constant of proportionality independent of $m$ and $n$.
The derivative of this with respect to the lowest-dimension
coupling $\tau$ must then be essentially the constant $r$ defined
in \instanton. Indeed, a calculation similar to \leadcl\ shows
that
\eqn\instZderII{
\partial_\tau Z_{m,n} \propto \int_{z_{m,n}}^{z_{m,n}+r_{m,n}} dz
= r_{m,n}
}
as expected. This provides a non-trivial check of the formula
\instZ\ for the instanton actions derived in \SeibergNM. It also
generalizes (and simplifies) the analysis of \KazakovDU, where
\instZderII\ was proven for the special case of the conformal
background.

In the conformal background, one can check that the instanton
actions $r_{m,n}$ are always real. However, in a general
background they will be complex. For instance, in the $(2,2m-1)$
models perturbed by the lowest-dimension operator, one can use the
formulas in \EynardSG\ to prove this explicitly for $m$
%%%
odd.   When
the $r_{m,n}$ are complex, the corresponding $(p,q)$ minimal
string theory is expected to be nonperturbatively consistent and
Borel summable. In these cases, the $r_{m,n}$ come in conjugate
pairs, so that even though they are complex, the total instanton
correction to the partition function is real.

\appendix{B}{A Brief Review of Stokes' Phenomenon}

In this appendix we will briefly review Stokes' phenomenon,
summarizing \Berry. Consider the following integral
 \eqn\inte{
 I(x) =
 \int_{\CC_0} ds\ e^{-{1\over \hbar} \CS(s,x)}
 }
where $\CS(s,x)$ is holomorphic in $s$, and $\CC_0$ is a contour in
the complex $s$-plane, chosen so that the integral exists and
admits an analytic continuation to some region of the complex
$x$-plane. We are interested in the $\hbar \to 0$ asymptotics.

%The first step is to extend $\CS$ to an analytic function of $s$
%and $x$ in some region around the real axes of these variables.
%
Since ${\partial \CS \over\partial \bar s}=0$, lines of constant
$\Im \CS$ are perpendicular to lines of constant $\Re \CS$; i.e.\
they are gradient lines of $\Re \CS$. We would like to deform the
contour $\CC_0$ in \inte\ to a steepest descent contour $\CC$ -- a
gradient line of $\Re \CS$ along which $\Im \CS$ is constant. (The
latter requirement prevents cancellation between different
non-saddle portions of the contour in the leading $\hbar\to 0$
approximation.) At a generic point such lines do not intersect.
However, the saddle points ${\partial \CS \over\partial s}=0$ are
characterized by having two intersecting steepest descent lines.

Since typically the different saddle points occur at different
values of $\Im \CS$, it is impossible to deform the contour
$\CC_0$ to a steepest descent contour (constant $\Im \CS$) $\CC$
passing through all of them.  However, if the steepest descent
contours $\CC_{1,2}$ through two different saddles labelled by $1$
and $2$ pass near each other, and have the proper asymptotic
behavior, we can deform $\CC_0$ as follows.  We deform it to a
steepest descent contour $\CC$ which starts close to $\CC_1$
passes near the saddle point $1$, then passes near the saddle
point $2$ and finishing close to $\CC_2$. Such a contour must be
compatible with the asymptotic behavior of the original contour
$\CC_0$. Alternatively, if $\CC_1$ and $\CC_2$ asymptote to each
other at infinity and $\Re \CS \to +\infty$ there, we can take
$\CC=\CC_1+\CC_2$ (see figure 1 and the example below).  This
makes it clear that the two saddles contribute to the integral.

Now let us vary the parameter $x$ in \inte\ and examine the
saddles and the contour $\CC$. There are two interesting things
that can happen. The first, more trivial phenomenon is when the
two saddles exchange dominance. This occurs across lines in the
complex $x$ plane called ``anti-Stokes lines," where the values of
$\Re\CS$ at the two saddles are the same. The second, more
interesting critical behavior happens across the ``Stokes lines,"
where the values of $\Im \CS$ at the two saddles are the same
and the   topology of $\CC_{1,2}$ changes. Beyond this
point the contour with the correct asymptotic behavior, or
equivalently a smooth deformation of the previous contour $\CC$,
does not pass through the two saddles but only through one of
them.  It is possible to find another contour which passes through
both of them, but it does not have the correct asymptotic
behavior. The exchange of dominance of two saddles and the abrupt
disappearance of the saddle-point contribution to the integral
$I(x)$ both contribute to Stokes' phenomenon. As mentioned in the
body of the paper, this is the phenomenon in which the analytic
continuation of the asymptotic expansion of a function does not
agree with the asymptotic expansion of the functions' analytic
continuation.

\medskip \ifig\figI{The steepest descent lines $\CC_{1,2}$ pass through
the saddles points $1$ and $2$.  The dotted line is the original
integration contour $\CC_0$.  For one value of $x$ the situation
is as in figure A, and $\CC_0$ can be replaced by
$\CC=\CC_1+\CC_2$ because the two contours $\CC_1$ and $\CC_2$
meet at an asymptotic infinity where the integrand vanishes. Then
the integral receives contributions from the two saddles. For
another value of $x$, as in figure B, the steepest descent contour
is given by $\CC=\CC_2$ alone, and so the integral receives a
contribution only from the saddle $2$. The transition occurs for
the values of $x$ for which $\Im \CS(1)=\Im \CS(2)$.}
    {\epsfxsize=0.6\hsize\epsfbox{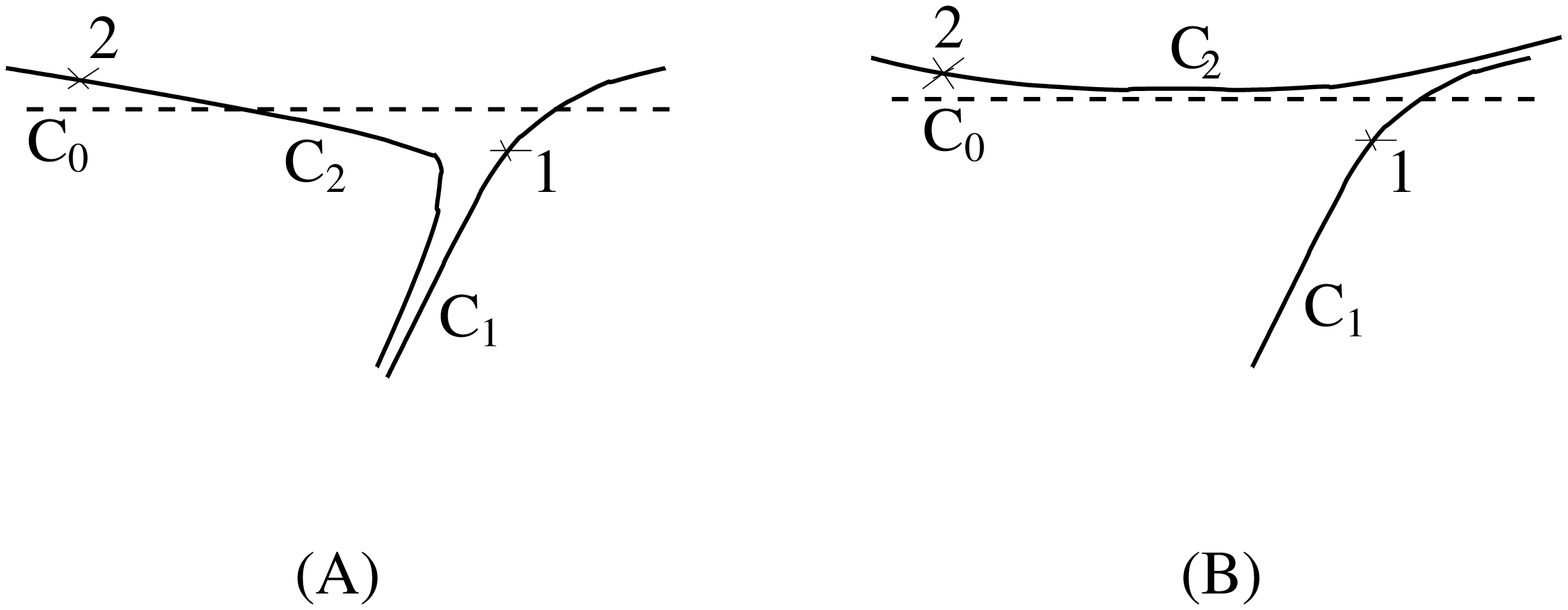}}

As an example, consider the Airy function
 \eqn\Air{\int_{-\infty}^{+\infty }ds e^{{i\over \hbar}({s^3 \over 3}
 + x s)}}
The behavior as $|s|\to \infty$ allows us to deform the contour to
start in the wedge ${2 \pi \over 3} \le \arg(s) \le  \pi $ and end
in the wedge $0 \le \arg(s) \le {\pi \over 3} $.  The two saddles
at $s=\pm \sqrt{-x} $ are as in figure 1. There is an anti-Stokes
line located on the negative $x$ axis. Here the two saddles are
purely imaginary (i.e. $\Re \CS=0$) and they exchange dominance.
One can also check that the lines $|\arg(x)|={2\pi\over3}$ are
Stokes lines. Thus, Figure A applies to ${2 \pi \over 3} \le |\arg
(x)| \le \pi$ and Figure B applies to $ 0 \le |\arg (x)| \le {2
\pi \over 3} $. As one crosses the Stokes lines starting from the
negative real axis, the dominant saddle ceases to contribute.

\appendix{C}{Numerical Analysis of $(p,q)=(2,5)$}

In this appendix we will analyze in detail the example of
$(p,q)=(2,5)$, using numerical methods where necessary. The
purpose of this analysis is mainly to verify that the lessons we
learned from the example of $(2,1)$ indeed carry over to more
complicated models.

To begin, we define the Lax operators to be
\eqn\LaxPQ{
Q=d^2-u(\tau),\qquad P = \sum_{k=0}^{2}t_{2k+3} Q^{k+1/2}_+,\qquad
d=\hbar\,\partial_\tau
}
This describes a perturbation around the $(2,5)$ multi-critical
point. Let us set $t_7=-8/5$ without loss of generality. Then the
string equation $[P,Q]=\hbar$ takes the form
\eqn\strmthr{
u^3+{3\over 4}t_5u^2-t_3u-\tau -{1\over4}\hbar^2\left(2u'^2+
4uu''+t_5u''\right) + {1\over10}\hbar^4 u^{(4)}=0
}
The Baker-Akhiezer function is determined by the differential
equations
\eqn\bakerex{
Q\psi = x\psi,\qquad P\psi=\hbar\partial_x\psi
}
together with the condition that $\psi$ is real and exponentially
decreasing as $x\to+\infty$.

Before we proceed to solve \strmthr\ and \bakerex\ numerically,
let us first discuss the classical limit $\hbar\to0$. At
$\hbar=0$, $P$ and $Q$ take the form
\eqn\LaxPQcl{
Q=d^2-u,\qquad P = -{8\over 5}d^5 + (4u+t_5)d^3 -
    (3u^2+ {3\over 2}t_5u - t_3)d
}
with $u(\tau)$ the solution to \strmthr\ with $\hbar=0$. Therefore
they lie on the Riemann surface described by the algebraic
equation
\eqn\Mpqcl{
P^2 ={1\over25}(Q + u)\left(8Q^2 - 4\left(u + {5\over 4}
t_5\right) Q + 3 u^2 + {5\over2}t_5 u - 5  t_3\right)^2
}
Here we see explicitly how the Riemann surface takes the form
\Mpqgreg\ for all values of the closed-string couplings
$\tau,t_3,t_5$. In particular, the Riemann surface always has a
branch point at $Q=-u$ and singularities at the other roots of the
RHS of \Mpqcl. Therefore the KdV flows, which change the values of
the closed string couplings, indeed correspond to
singularity-preserving deformations of the Riemann surface.

Now let us discuss the numerical solution of the string equation
\strmthr\ and the Baker-Akhiezer equations \bakerex. The string
equation for perturbations around the $(p,q)=(2,5)$ critical point
was solved numerically in \refs{\BrezinVD,\DouglasXV}. Here we
will repeat the analysis of \refs{\BrezinVD,\DouglasXV} to obtain
the specific heat $u(\tau)$ for various values of $\hbar$. We will
then take the analysis one step further by numerically solving
\bakerex\ for the Baker-Akhiezer function. For simplicity, let us
limit ourselves to the conformal background perturbed by the
lowest-dimension operator. The conformal background corresponds to
\eqn\confbg{
\tau=0,\qquad t_3=1,\qquad t_5=0
}

Up to a trivial shift of $u$ and $\tau$ this is identical to the
setup considered in \DouglasXV. To see that this is the conformal
background, simply substitute \confbg\ into the formula \Mpqcl\
for the Riemann surface. Since the string equation \strmthr\ is
solved by $u(\tau=0)=1$ (modulo a discrete choice for the root of
the cubic polynomial) the curve becomes
\eqn\Mpqclconf{
y^2={4\over25}(x+1)(4x^2-2x-1)^2
}
which is indeed the same as $T_2(y)=T_5(x)$ after a rescaling of
$y$.

Shown in figure 2 is the specific heat $u(\tau)$ versus $\tau$ for
various values of $\hbar$. At large $|\tau|$ the specific heat
asymptotes to the classical solution $u_{\rm cl}(\tau)\sim {\rm
sign(\tau)}|\tau|^{1/3}$. Meanwhile, at small $|\tau|$ the
specific heat oscillates faster and faster as $\hbar$ is
decreased, since here the function is trying increasingly hard to
interpolate smoothly between the classical discontinuity $u_{\rm
cl}(\tau=0)=\pm 1$ at $\tau=0$. Evidently, the classical limit of
$u(\tau)$ is not well-defined for small $|\tau|$, although the
quantum answer is smooth.

\medskip \ifig\figII{The specific heat $u(\tau)$ as a function of
the lowest-dimension coupling $\tau$, for $\hbar=1,0.5,0.3$. These
plots were obtained by numerically solving the string equation
\strmthr\ in the conformal background \confbg.}
    {\epsfxsize=0.8\hsize\epsfbox{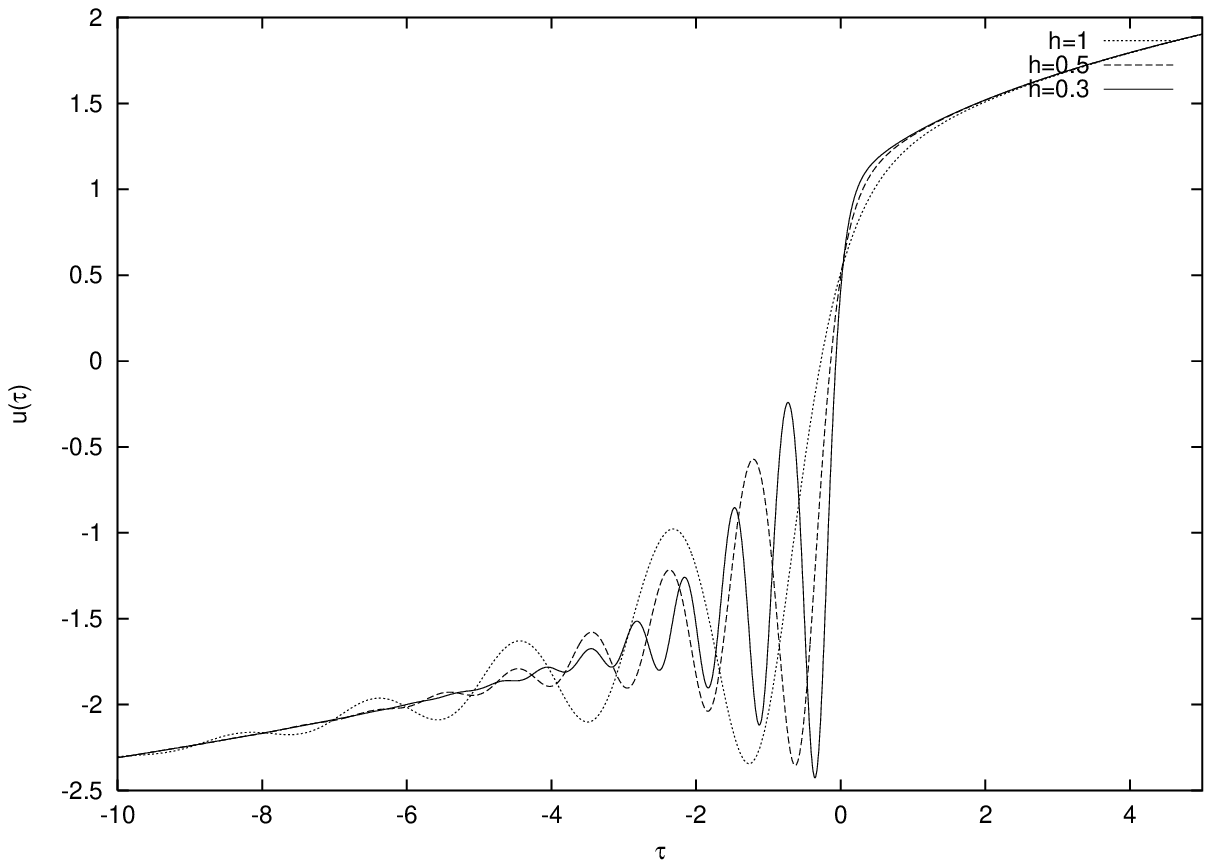}}

Figure 3 contains a plot of the Baker-Akhiezer function
$\psi(x,t)$, again for various values of $\hbar$. (The different
solutions have been rescaled in order to aid the presentation.)
{}From the figure, it is clear that $\psi(x,t)$ is decreasing at
large positive $x$, while it is oscillatory for $x<-1$. Also, the
function is clearly always smooth and real-valued. The bump at $x
\approx {1+\sqrt{5}\over 4}$ in figure 3 corresponds to the
location of the $(1,2)$ ZZ brane, while the trough at $x\approx
{1-\sqrt{5}\over4}$ is the location of the $(1,1)$ ZZ brane. As
$\hbar$ decreases, the oscillations at $x<-1$ become faster, and
the bump at $x\approx {1+\sqrt{5}\over 4}$ becomes more
well-defined. This behavior is all qualitatively consistent
with the leading-order WKB approximation \eqn\psicl{ \psi_{\rm
cl}(x,t) \approx \cases{ (-1-x)^{-1/4} e^{\int_{-1}^{x}
y\,dx'/\hbar} & $x>-1$\cr 2(x+1)^{-1/4}\sin\left({\pi\over
4}-{i\over\hbar}\int_{-1}^{x} y\,dx'\right) & $x<-1$} } where
$y=y(x)$ is given by \Mpqclconf.

\medskip \ifig\figIII{The Baker-Akhiezer function $\psi(x,t)$ versus the boundary
cosmological constant $x$, for $\hbar=1,0.5,0.3$. The plots have
been rescaled for the different values of $\hbar$, so as to
improve the presentation.}
    {\epsfxsize=0.7\hsize\epsfbox{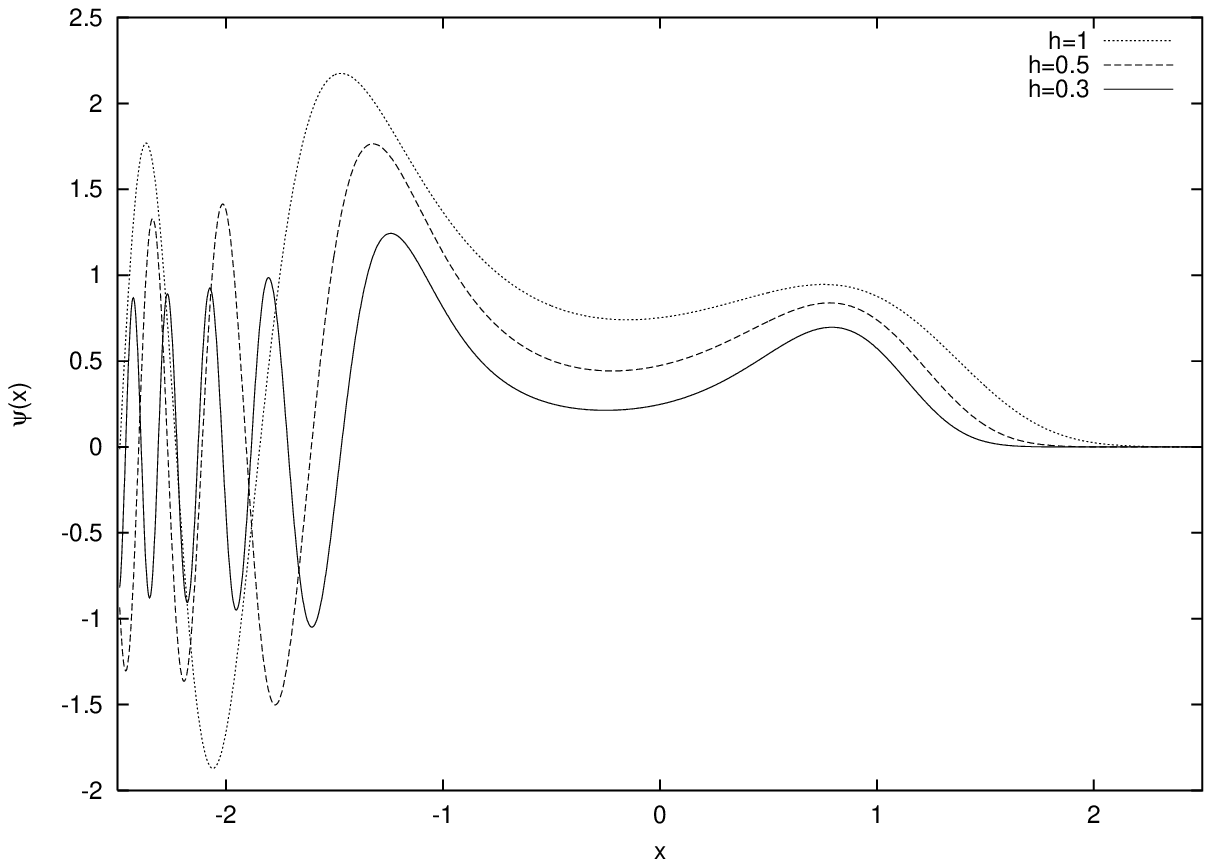}}

A more quantitative comparison between the WKB approximation and
the exact answer is shown in figure 4, $\hbar=0.3$. We see that
they are in excellent agreement, except for a small region around
$x=-1$ where we expect the WKB approximation to break down anyway.

It should  be clear from the discussion that these numerical
results confirm many of the general arguments in the text
regarding the properties of the Baker-Akhiezer function
$\psi(x,t)$. Let us just mention a few. First, $\psi(x,t)$
obviously exhibits Stokes' phenomenon: the analytic continuation
of the asymptotics \psicl\ away from large positive $x$, where
$\psi(x,t)$ is exponentially decreasing, leads to the wrong answer
for $x<-1$, where $\psi(x,t)$ is oscillatory. Second, notice that
the analytic continuation of the WKB approximation from large
positive $x$ is accurate up until $x\approx-1$. The failure of the
analytic continuation of the WKB approximation beyond $x=-1$ is
due to the level crossing phenomenon, which results in the
oscillatory behavior of $\psi(x,t)$. These facts agree well with
the general discussion in section 2. Finally, note that the
Baker-Akhiezer function is exponentially {\it decreasing} at large
positive $x$. From section 5, we know that this is the expected
behavior for the nonperturbatively consistent $(2,5)$ model.

\medskip \ifig\figIV{A comparison of exact Baker-Akhiezer function and its
leading-order WKB approximation, for $\hbar=0.3$. The two are
clearly in excellent agreement, except in a small region around
$x=-1$ where the WKB approximation is expected to break down.
}
    {\epsfxsize=0.6\hsize\epsfbox{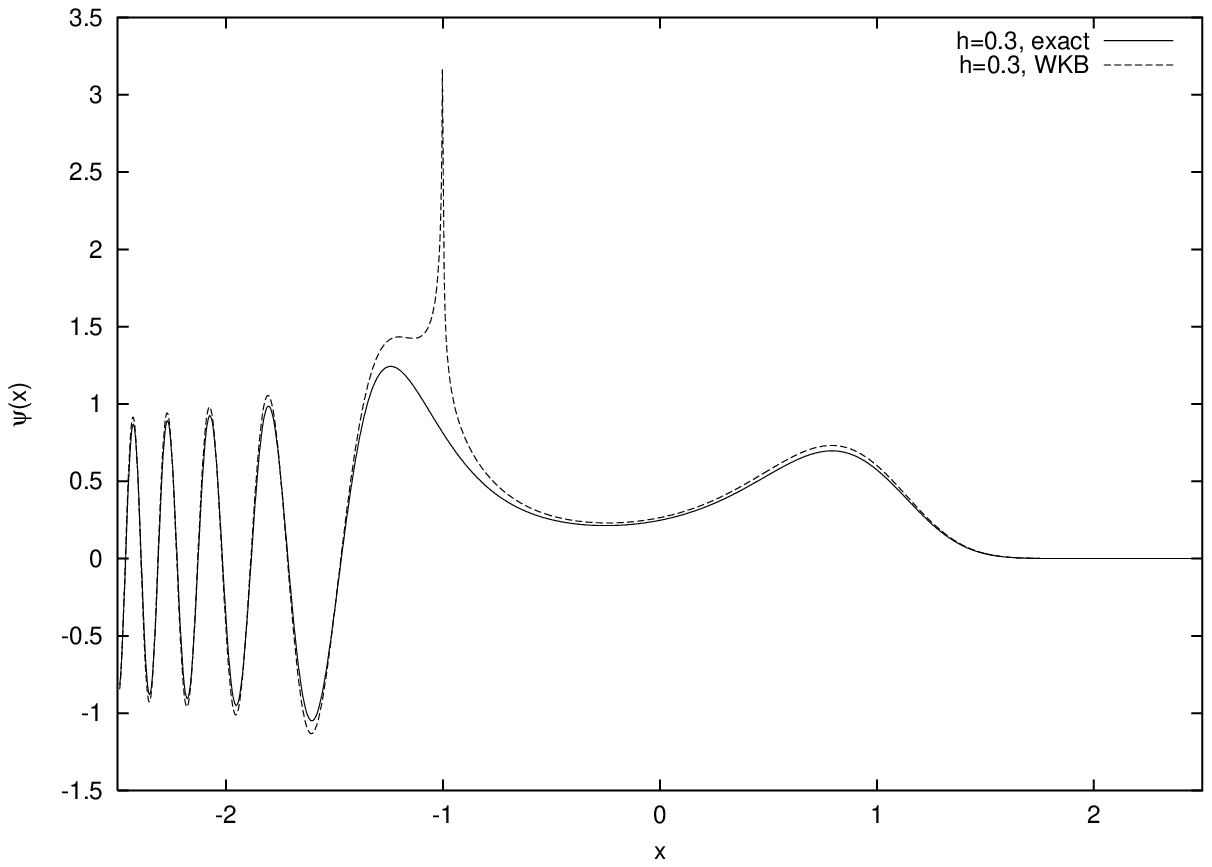}}

\listrefs
\end